\documentclass[12pt]{article}
\pdfoutput=1 

\usepackage{jheppub} 
                                          
\usepackage{placeins}
\usepackage{tabularx}
\usepackage{amsmath}

\newcommand{\rmii}[1]{{\mbox{\tiny\rm{#1}}}}
\newcommand{\beq}{\begin{equation}}
\newcommand{\eeq}{\end{equation}}
\newcommand{\bea}{\begin{align}}
\newcommand{\eea}{\end{align}}
\newcommand{\beas}{\begin{align*}}
\newcommand{\eeas}{\end{align*}}

\newcommand{\bp}{{\bf p}}

\newcommand{\mD}{m_\rmii{D}}

\newcommand{\Tint}[1]{{\hbox{$\sum$}\!\!\!\!\!\!\!\int\,}_{\!\!\!\!\raise-0.9ex\hbox{$\scriptstyle{#1}$}}}
\renewcommand{\Re}{\rm Re}
\renewcommand{\Im}{\rm Im}

\definecolor{dunkelgrau}{rgb}{0.43,0.43,0.43}

\title{\bf\large In-medium P-wave quarkonium from the complex lattice QCD potential}

\author[a]{Yannis Burnier,}
\author[b]{Olaf Kaczmarek}
\author[c]{and Alexander Rothkopf}

\affiliation[a]{Institute of Theoretical Physics, EPFL, CH-1015 Lausanne, Switzerland}
\affiliation[b]{Fakult\"at f\"ur Physik, Universit\"at Bielefeld, D-33615 Bielefeld, Germany}
\affiliation[c]{Institute for Theoretical Physics,  Heidelberg
  University, Philosophenweg 16, 69120 Heidelberg, Germany}

\emailAdd{yannis.burnier@epfl.ch}
\emailAdd{okacz@physik.uni-bielefeld.de}
\emailAdd{rothkopf@thphys.uni-heidelberg.de}

\date{\today}

\abstract{We extend our lattice QCD potential based study \href{http://dx.doi.org/10.1007/JHEP12(2015)101}{[JHEP 1512 (2015) 101]} of the in-medium properties of heavy quark bound states to P-wave bottomonium and charmonium. Similar to the behavior found in the S-wave channel their spectra show a characteristic broadening, as well as mass shifts to lower energy with increasing temperature.  In contrast to the S-wave states, finite angular momentum leads to the survival of spectral peaks even at temperatures, where the continuum threshold reaches below the bound state remnant mass. We elaborate on the ensuing challenges in defining quarkonium dissolution, present estimates of melting temperatures for the spin averaged $\chi_b$ and $\chi_c$ states and contrast the findings to recent direct lattice NRQCD studies of P-wave quarkonium. As an application to heavy-ion collisions we estimate the contribution of feed down to S-wave quarkonium through the P-wave states after freezeout.}

\begin{document}

\maketitle


\flushbottom

\FloatBarrier

\section{Introduction}

With the recent discovery of the third spin triplet $\chi_b(3P)$ and its radiative decays to the well established $\Upsilon(1S,2S,3S)$ states in $pp$ collisions at the LHC \cite{Aaij:2014caa}, a complete picture of the S- and P-wave quarkonium bound states below the open heavy flavor threshold has emerged. The high precision determination of its mass \cite{Aad:2011ih,Abazov:2012gh,LHCb:2012mja} as well as its contribution to feed-down complements the comprehensive body of studies performed on the properties and decay channels of both charmonium at CDF \cite{Abulencia:2007bra}, HERA-B \cite{Abt:2008ed}, LHCb \cite{LHCb:2012ac,LHCb:2012af,Aaij:2013dja} and BESIII \cite{Ablikim:2011kv}, as well as those for the bottomonium family at CDF \cite{Affolder:1999wm}, ATLAS \cite{Aad:2011ih}, CMS \cite{Khachatryan:2014ofa} and LHCb\cite{Aaij:2012se,LHCb:2012mja}. On the side of theory the maturation of effective field theory frameworks, such as NRQCD \cite{Brambilla:2004jw}, systematically derived from Quantum Chromo Dynamics (QCD), have made it possible to compute many of the properties of heavy quarkonium in concert either with perturbation theory or lattice QCD simulations. And indeed both the $J/\Psi$ polarization puzzle at Tevatron \cite{Brambilla:2010cs,Braaten:1999qk,Affolder:2000nn} and the challenges associated with reproducing polarization effects in hadroproduction of $\Upsilon$ states at LHC have revealed that a thorough understanding of finite angular momentum bound states in vacuum is essential for charmonium \cite{Chao:2012iv,Gong:2012ug} as well as bottomonium \cite{Gong:2013qka,Feng:2015wka}. The experimental studies showed on the one hand that the feed-down from $\chi_b(3P)$ to $\Upsilon(3S)$, formerly disregarded in theory, can be sizable, while on the other hand the theory computations themselves concluded that e.g. without a feed-down contribution of more than $30\%$ to $\Upsilon(2S)$ they are unable to reproduce the measured yields at LHC.
                                          
In contrast to our detailed knowledge of quarkonium states in vacuum, their study at finite temperature has only recently entered an era of high precision. Motivated originally by the work of Matsui and Satz \cite{Matsui:1986dk} that proposed the melting of charmonium as clear signal of the formation of a deconfined quark-gluon plasma in relativistic heavy-ion collisions, experimental efforts to measure in-medium quarkonium properties have seen significant progress at RHIC and LHC.  The observation of a strong relative suppression of bottomonium S-wave excited states \cite{Chatrchyan:2011pe,Chatrchyan:2012lxa} and the replenishment of yields of $J/\Psi$ \cite{Abelev:2012rv}, the charmonium vector channel ground state, to name a few, have been highlights in this regard. Results on finite angular momentum states have so far not yet been obtained in heavy-ion collisions. Their influence through feed-down on the abundances of S-wave states in vacuum however promises that their role in-medium will also be non-negligible. On the one hand their larger spatial extend compared to S-wave states is expected to make them more susceptible to thermal fluctuations, while the presence of finite angular momentum can provide a stabilizing effect. Therefore a quantitative investigation of their in-medium behavior appears timely.

In the language of field theory the presence and properties of bound states in vacuum and in-medium can be deduced from the spectral function of an appropriately projected current-current correlator. Bound states available to a $Q\bar{Q}$ pair correspond to skewed Breit-Wigner type peaks located at certain frequencies and which can exhibit a finite width. The position of a peak can be understood as the mass of the state, while the width corresponds to its inverse lifetime. This width contains information on both the quarkonium state transitioning from and to a different state in the same channel, as well as the possibility to leave the channel under consideration, e.g. changing from a color singlet to an octet configuration at finite temperature. The computation of such spectral functions for bottomonium and charmonium states close to the deconfinement transition has been pursued by several groups in recent years using three separate non-perturbative lattice QCD methods.

The first two approaches ( \cite{Asakawa:2003re,Datta:2003ww,Jakovac:2006sf,Iida:2006mv,Aarts:2007pk,Ohno:2011zc,Ding:2012sp,Borsanyi:2014vka,Ohno:2014uga} and \cite{Aarts:2014cda,Aarts:2012ka,Skullerud:2014sla,Kim:2014iga,Kim:2014nda,Kim:2015csj,Kim:2015rdi}) compute the quarkonium spectra from lattice QCD correlators in Euclidean time, where the propagation of a kinetically equilibrated heavy quark pair in a thermal medium is simulated on a finite space-time grid. In the former studies, bottom or charm are treated as fully relativistic fields, i.e. one does not introduce any further approximations to the heavy quarks. This approach however requires very finely spaced lattices to resolve the quarkonium states, which currently limits its application to simulations without dynamical light fermion flavors, i.e. quenched QCD.

In the latter studies a non-relativistic QCD (NRQCD) formulation is considered instead. Lattice NRQCD \cite{Lepage:1987gg,Lepage:1992tx} relies on an expansion of the QCD Lagrangian in terms of increasing powers of the heavy quark velocity and incorporates additional radiative corrections through the inclusion of QCD Wilson coefficients, often determined perturbatively.  While at first sight it appears to suffer from the unavailability of a naive continuum limit, the current generation NRQCD codes incorporate up to ${\cal O}(v^6)$ corrections \cite{Meinel:2010pv} and in combination with dynamical simulations for light flavors have been shown to reproduce the vacuum properties of charmonium and bottomonium to high precision. This made it even possible to predict e.g the $\eta_b(2S)$ mass \cite{Dowdall:2011wh} before its experimental discovery. The application of lattice NRQCD to finite temperature is well established, at least for bottomonium at the temperatures currently reached in heavy-ion collisions. 

To obtain spectral functions from lattice simulations carried out in Euclidean time one has to perform an analytic continuation using a finite number of stochastically approximated correlator points. This constitutes an ill-defined inverse problem. Two methods based on Bayesian inference are currently in use. The well established Maximum Entropy Method (MEM) \cite{Asakawa:2000tr, Rothkopf:2012vv,Rothkopf:2011ef} tends to deliver smooth reconstructed spectra but is susceptible to under-predicting peaked features. The recently proposed Bayesian Reconstruction (BR) method \cite{Burnier:2013nla,Burnier:2013esa} on the other hand, which allows higher resolution in the determination of peaked structures can introduce nonphysical ringing in the reconstruction. Since a collection of intricate features, such as bound state peaks, the open heavy flavor threshold and eventually the lattice cutoff is present in the actual spectrum, its reliable reconstruction from Euclidean data is highly challenging. These difficulties are particularly apparent when using Euclidean data from simulations with realistic light quarks, where currently only a very small number of data points $\sim{\cal O}(10)$ are available \cite{Bazavov:2011nk,Borsanyi:2013bia,Bazavov:2014pvz}. Two groups have recently investigated bottomonium states using Bayesian reconstruction methods in lattice NRQCD at finite temperature \cite{Aarts:2014cda,Kim:2014iga}. They reported in mutual agreement that a spectral feature corresponding to the $\Upsilon(1S)$ state survives far into the high temperature QGP phase up to at least $T=1.6T_C$. The situation for P-wave bottomonium is more ambiguous, since depending on the deployed method the results differ. For $\chi_b(1P)$ the MEM based analyses found vanishing peak structures shortly above the deconfinement transition around $T_{\rm melt}^{\rm MEM}\approx 1.27T_C$ while reconstructions based on the BR method revealed weakly peaked bound-state features even up to $1.6T_C$. In this study we provide an independent analysis of the spectral structures of P-wave bottomonium that allows us to proceed towards a resolution of this issue.

The third available route to in-medium quarkonium spectra, which is taken here, is based on a lattice QCD determination of the complex valued real time potential and amounts to using a non-relativistic potential description known as pNRQCD \cite{Brambilla:1999xf}. This EFT allows one to systematically define an in-medium potential that in turn can be computed using either resummed perturbation theory \cite{Laine:2006ns,Brambilla:2008cx} at high temperature or non-perturbatively from lattice QCD \cite{Rothkopf:2011db,Burnier:2012az,Burnier:2014ssa}. Since pNRQCD is derived from NRQCD, it too can be systematically improved going beyond the static inter-quark potential, amending it with e.g. velocity and spin dependent corrections \cite{Brambilla:2000gk,Pineda:2000sz}.  In vacuum the values of the purely real static $T=0$ potential have long been known non-perturbatively, together with their corrections up to ${\cal O}(v^2)$ \cite{Koma:2006si,Koma:2006fw}. At finite temperature it has only recently become possible to define and extract the inter-quark potential from the Euclidean Wilson loop by using Bayesian inference methods. The extraction of higher order corrections for this quantity is a topic of active research. 

In pNRQCD the quarkonium unequal time correlator (not the two-body wavefunction itself) is propagated via a Schr\"odinger equation hosting the lattice QCD in-medium potential. The real and imaginary component of the potential encode the interaction of the $Q\bar{Q}$ state with the surrounding thermal bath.  As the quarkonium correlator is related to its spectral function via a Fourier transform we solve its Schr\"odinger equation directly in frequency space \cite{Burnier:2007qm}. The computation is performed directly in Minkowski space-time, without the need for an analytic continuation at this step and the resulting spectra are devoid of the resolution artifacts encountered when reconstructing the spectrum from Euclidean correlators. The only time we require a Bayesian reconstruction is at the step of extracting the values of the in-medium potential \cite{Rothkopf:2011db}. Since the functional form of the corresponding spectrum is much simpler than the two-body $Q\bar{Q}$ spectrum itself \cite{Burnier:2012az}, it can be robustly determined with currently available methods and simulation data sets \cite{Burnier:2014ssa}.\footnote{In fact the spectral function of the potential contains one dominant peak and falls off quickly at large and small frequencies so that the corresponding Euclidean correlator, i.e. the Coulomb gauge Wilson lines, is finite at $\tau=0$ also in the continuum\cite{Burnier:2013fca,Burnier:2009bk,Berwein:2012mw}.} Our calculation of in-medium P-wave properties uses the latest determination of the static complex in-medium potential from dynamical lattice QCD as in our previous S-wave study in Ref. \cite{Burnier:2015tda}.

In the subsequent section \ref{Sec:QuarkMelt} we will discuss the challenges of defining what constitutes melting in particular for a P-wave state and more generally how to relate spectral peak features to the presence or absence of in-medium bound states. Section \ref{Sec:NumRes} first summarizes the properties of the in-medium potential we deploy, before presenting the resulting finite temperature spectra for P-wave bottomonium and charmonium and their melting temperatures. We discuss our findings and compare to those of recent direct lattice NRQCD spectral computation. The in-medium modification of the spectral features is put in the context of feed-down to the S-wave states in heavy-ion collisions in section \ref{Sec:HeavyIon} before we summarize our study and provide concluding remarks in section \ref{Sec:Concl}.

\FloatBarrier

\section{Quarkonium melting}
\label{Sec:QuarkMelt}

Ever since the classic works of Matsui and Satz \cite{Matsui:1986dk}, quarkonium melting has been a central topic in the study of the Quark-Gluon Plasma, nowadays accessible at heavy-ion collider facilities. The original idea envisaged a situation, where a (kinetically) thermalized quark-antiquark pair forms an in-medium Eigenstate whose wavefunction propagates in time only with a simple phase. In such a static scenario bound states either exist or have melted. Earlier works in this field \cite{Nadkarni:1986cz,Nadkarni:1986as,Wong:2004zr,Kaczmarek:2005ui,Kaczmarek:2005gi,Kaczmarek:2007pb,Satz:2008zc}, which have shaped our intuition, used purely real model potentials, in which the determination of what constitutes a bound state could be readily read off from the long-distance behavior of the corresponding wavefunction or by computing the binding energies. Note that in the presence of confinement, where the remnant of a linear contribution of the inter-quark potential only gradually decreases with temperature, the binding energy is defined from the difference between the energy of the state to the continuum threshold, which in pNRQCD is related to the large distance behavior of the real part of the potential. Contrary to the Coulomb case this threshold moves to lower energies with increasing temperatures.

With the maturation of effective field theory methods, such as NRQCD and pNRQCD it became possible to leave the realm of model potentials and to compute the proper in-medium inter-quark potential both with perturbative and lattice QCD techniques. It was found that the heavy-quark potential actually takes on complex values. The appearance of an imaginary part is related to the fact that the object described by the corresponding Sch\"odinger equation is not the wavefunction of the quarkonium state but the medium averaged unequal time meson correlator. In the language of pNRQCD this correlator can be understood as the medium average of the product of the $Q\bar{Q}$ wavefunction at initial time multiplied with the wavefunction of the system after evolving to real-time $t$. A decay in the correlator corresponding to a finite $\Im[V]$ hence represents the decoherence of the initial wavefunction through the kicks it receives from the surrounding thermal medium. Note that it is not related to the annihilation of the heavy quark-antiquark pair itself. 

The meson correlator is connected instead to the quarkonium spectral function of the system via a Fourier transform. The imaginary part of the inter-quark potential induces broadening of peak structures that originally at $T=0$ represent the well defined vacuum bound states. The presence of a width tells us that we may not consider the quarkonium system as static, but instead it is highly dynamical. The inverse of the spectral width corresponds to the lifetime of the in-medium excitation that a particular peak structure represents. I.e. even if we start with the system in a particular in-medium state there exists a non-negligible probability for it to transition to either another neighboring state within the same channel or it may leave the channel all together, e.g. if a color singlet state converts to a color octet. In a potential based description of the wavefunction evolution the former may be described by a stochastic potential \cite{Akamatsu:2011se,Rothkopf:2013kya,Akamatsu:2014qsa,Akamatsu:2015kaa}, while the loss channel would require an actual imaginary part of the wavefunction potential. 

Since we do not have direct access to the wavefunction of the two-body system and the concept of a static in-medium Eigenstates is not applicable, any statement about quarkonium survival or melting needs to be defined based on spectral features. The most popular criterium, proposed in Ref.\cite{Laine:2006ns}, is to compare the spectral width of a state with its in-medium binding energy. If the two become of the same size we may consider the state as unstable against transitioning, declaring it melted. This does not mean that in the spectral function no peak structure is visible anymore. On the contrary, when investigating S-wave quarkonium states in \cite{Burnier:2015tda} we found that the remnant of the former bound state peak can survive to much higher temperatures after melting, according to the above criterium, has occurred. I.e. the influence of a state can e.g. survive as a form of threshold enhancement which in turn can give corrections to purely thermal estimates of quarkonium production.

The above definition of melting is meaningful in the S-wave channel due to the fact that the continuum threshold always remained above the bound state remnants mass. I.e. while the continuum moves to lower energies with increasing temperature it will eventually also push the bound state peaks to lower masses. Hence the spectral width of the in-medium excitation and its binding energy can be clearly identified at least until they agree in size. 

Note that in a heavy-ion collision we do not measure the in-medium states directly but instead vacuum states that have formed after the medium has reached freezeout and entered the hadronic phase. Thus even if there are no well defined features present in an in-medium spectral function, corresponding to an equilibrated quark-gluon plasma at high temperature, it does not mean that experiment will not eventually measure a finite yield of quarkonium states. At the transition to the confined phase individual $Q$ and $\bar{Q}$'s will have to form colorless hadrons, most of which end up as $D$ or $B$ mesons. However if the number of heavy quarks produced in the early stages of a heavy-ion collision is large enough there exists a non-negligible probability for them to come together as quarkonium too. How to describe this process of recombination in the language of quantum field theory is an open question. Here we attempt to do so by assigning the area under the in-medium spectrum, which lies close to the mass of a vacuum state, to the amplitude of the bound state delta peak at $T=0$. A similar strategy underlies e.g. phenomenological recombination approaches, such as the statistical model of hadronization, where even though one assumes full melting of all bound states in the medium, a sizable number of charmonium ground and excited states is produced after freezeout.

\FloatBarrier

\section{P-wave Quarkonium spectra}
\label{Sec:NumRes}

\subsection*{Phenomenological setup}

\begin{figure}
\centering
 \includegraphics[scale=0.35, trim= 0 1.5cm 0 1cm, clip=true ]{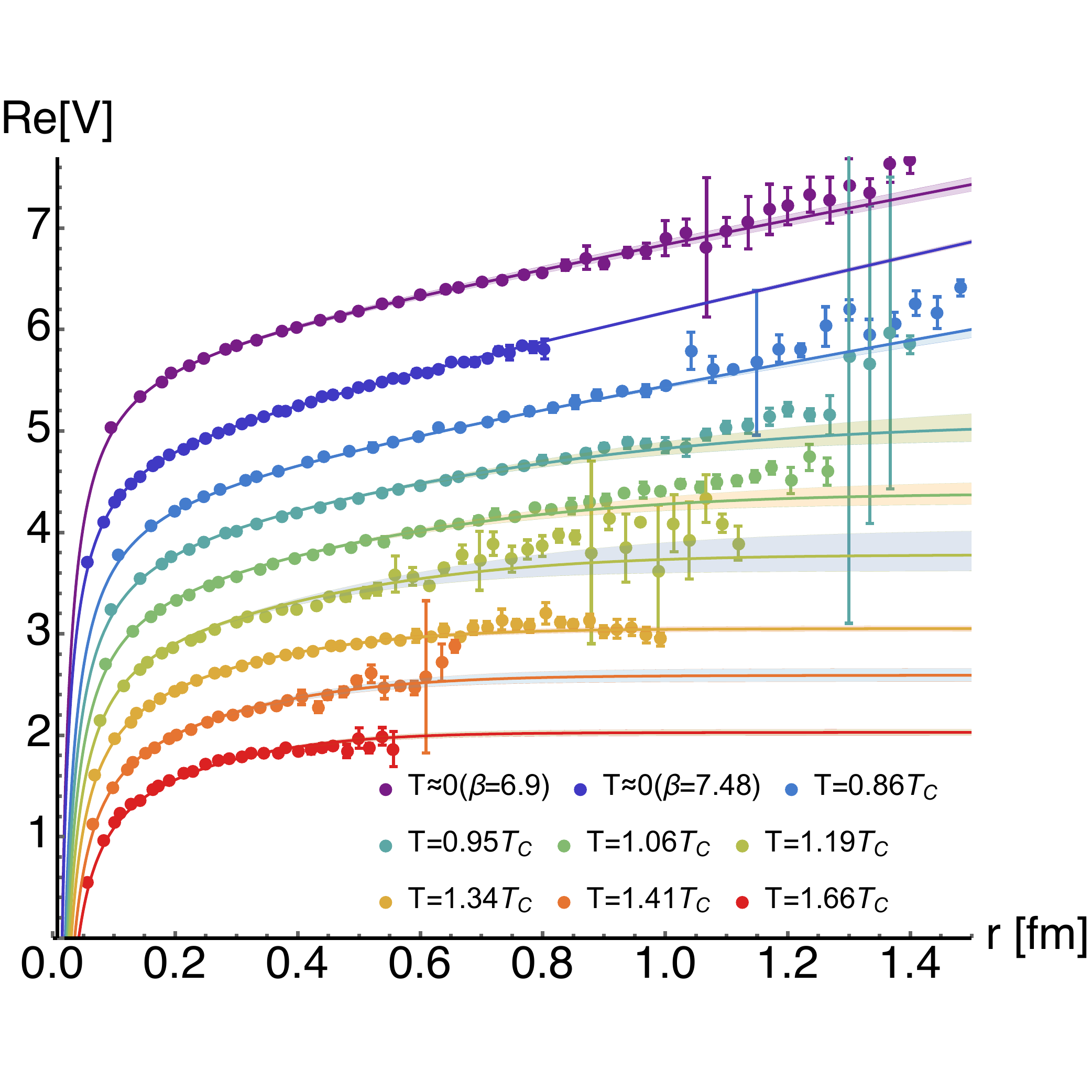}\hspace{0.2cm}
 \includegraphics[scale=0.35, trim= 0 1.5cm 0 1cm, clip=true ]{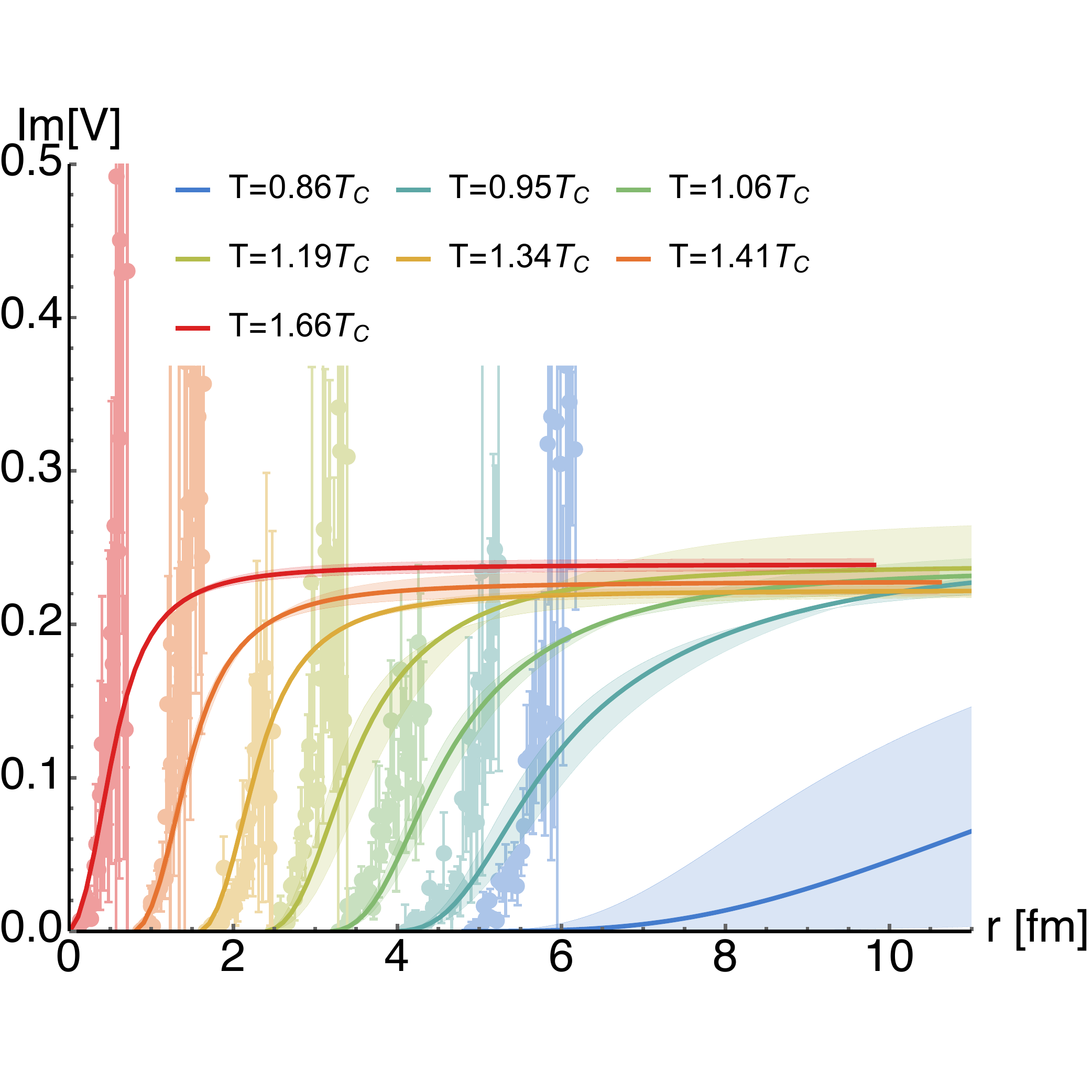}
 \caption{(left) Real-part of the static in-medium heavy quark potential in $N_f=2+1$ dynamical lattice QCD (points) and an analytic parametrization based on a single temperature dependent fit parameter $m_D$, identifies previously as a Debye mass. (solid line). Error bands denote changes from varying the value of $m_D$ within its fit uncertainty. (right) prediction of the imaginary part of the potential (solid curves), together with the tentative values (light points) extracted from the dynamical $N_f=2+1$ lattice QCD with $N_\tau=12$.}\label{Fig:ReImPot}
\end{figure}

In this study we will adopt the same potential based approach as in Ref.\cite{Burnier:2015tda}, which for completeness is summarized here briefly. Our starting point is a Cornell type $T=0$ potential with three parameters
\begin{align}
V^{\rm T=0}(r)=-\frac{\alpha_s}{r} +\sigma r +c,\label{Eq:Cornell}
\end{align}
the strong coupling $\alpha_s$, the string tension $\sigma$ and an arbitrary constant shift $c$. At first we fit the values of these vacuum parameters using the numerically extracted potential on $T\approx0$ configurations in $N_f=2+1$ dynamical lattice QCD provided by the HotQCD collaboration \cite{Bazavov:2011nk,Bazavov:2014pvz}. As can be seen by the two top most solid curves on the left in Fig.~\ref{Fig:ReImPot} this simple parametrization allows to reproduce the $T=0$ lattice data extremely well. Through the use of the extended Gauss Law ansatz introduced in \cite{Burnier:2015nsa} we obtain an analytic parametrization of the corresponding complex in-medium potential. It depends on a single temperature dependent parameter $m_D$ that has been proposed as definition of the Debye mass. 

\begin{table}[t!]
\centering
\begin{tabularx}{15.5cm}{ |>{\centering}m{1cm} | >{\centering}m{1.6cm}| >{\centering}m{1.6cm} | >{\centering}m{1.6cm} | >{\centering}m{1.6cm} | >{\centering}m{1.6cm} | >{\centering}m{1.6cm} | X | }
\hline
	$T/T_C$ & 0.86 & 0.95 & 1.06 & 1.19 & 1.34 & 1.41 & 1.66 \\ \hline
	$\frac{m_D}{\sqrt{\sigma}}$ & 0.01(3) & 0.25(8) & 0.39(8) & 0.53(21) & 0.96(5) & 0.99(13) & 1.27(8) \\ \hline
	$\frac{m_D}{T}$ & 0.04(10) & 0.72(22) & 1.03(22)&1.28(49)&2.07(11)&2.05(27)&2.29(14)  \\ \hline
\end{tabularx}
\caption{Debye masses from Ref.\cite{Burnier:2015tda} entering our study. For use in phenomenology, a continuum corrected $m_D$ may be obtained from the ratio $\mD/\sqrt{\sigma(\beta)}$ shown here, through a multiplication with the continuum value of $\sigma$.}\label{Tab:DebMass}
\end{table}

\begin{figure}[t!]
\centering
  \includegraphics[scale=0.35, trim= 0 2.5cm 0 1cm, clip=true]{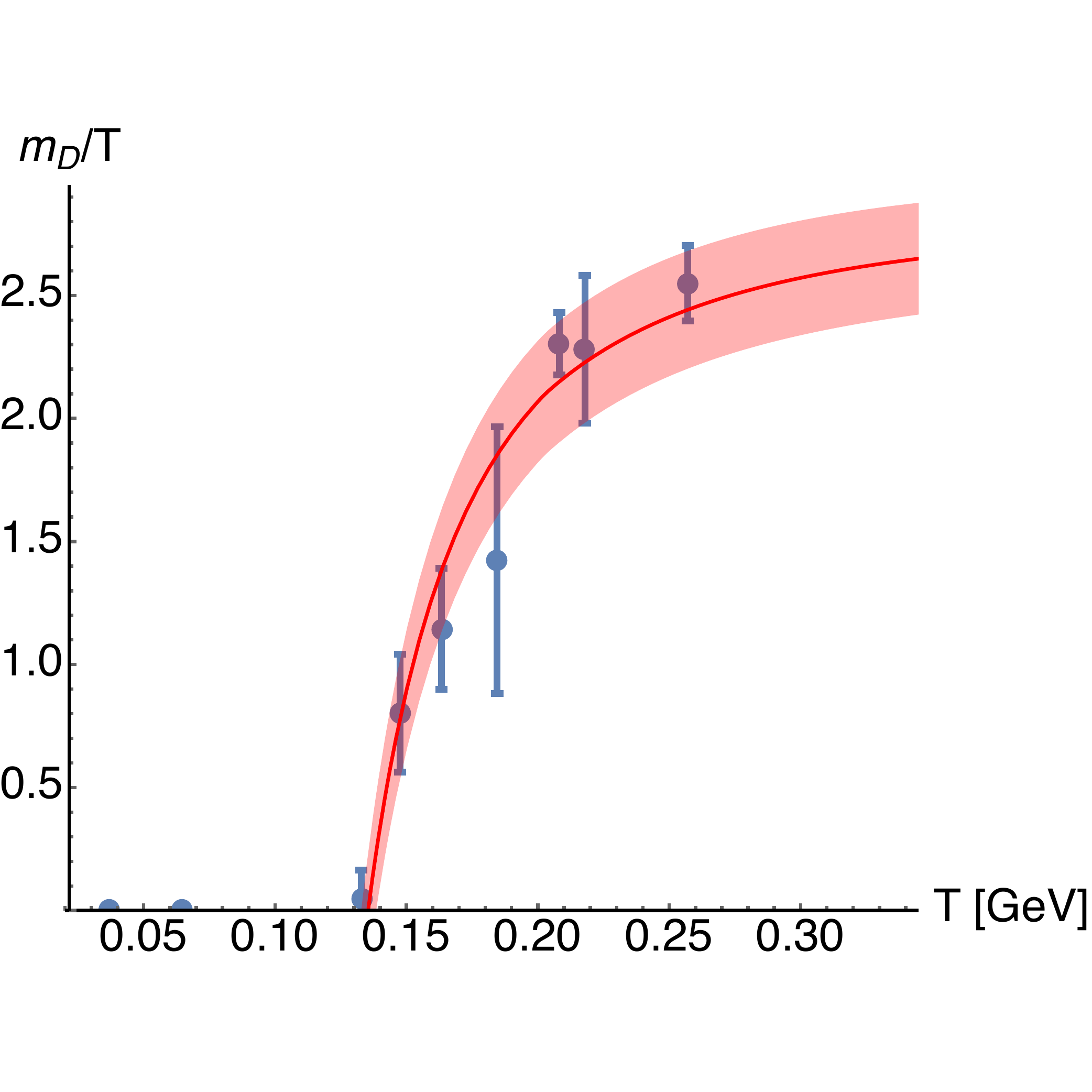}
 \caption{(left, blue points) The normalized Debye mass ($m_D/T$) from a fit of the extended Gauss-Law analytic parametrization to the real-part of the lattice QCD based in-medium heavy quark potential. These values encode the complete in-medium information entering in the computation of the P-wave spectra below. The red solid curve shows our interpolation of the temperature dependence of $m_D$, based fit function that goes over to the perturbative HTL behavior at large temperatures, while allowing for non-perturbative deviations from the weak-coupling temperature dependence close to the phase transition.}\label{Fig:mDFits}
\end{figure} 
\begin{table}
\centering
\begin{tabularx}{14cm}{ |>{\centering}m{3.7cm} | >{\centering}m{1.5cm}| >{\centering}m{1.5cm}| >{\centering}m{1.5cm} || >{\centering}m{1.5cm}  | X | }
\hline
 states    &  $\chi_b(1P)$ & $\chi_b(2P)$ & $\chi_b(3P)$ & $\chi_c(1P)$ & $\chi_c(2P)$ \\ \hline\hline
 $m$ [GeV] &  9.92597 & 10.269 & 10.538 & 3.5089 & 3.7918\\ \hline
 $m^{\rm \scriptscriptstyle PD
 G}$ [GeV]&   9.88814 & 10.252 & 10.534 & 3.4939 & 3.9228\\ \hline
 $\langle r \rangle$  [GeV$^{-1}$]& 2.435& 3.898& 5.586 & 4.136 & 25.42 \\ \hline
 $\langle r \rangle$ [fm] & 0.4797& 0.7679& 1.100& 0.814824 & 5.00813\\ \hline
 $\bar{m}_{B\bar{B}\,{\rm or}\,D\bar{D}}^{\rm \scriptscriptstyle PDG}-m$[GeV]&  0.633& 0.29 & 0.02& 0.227 & -0.056\\ \hline
\end{tabularx}
\caption{The masses, mean radii and distances to the open heavy flavor threshold for bottomonium (column 2-4) and charmonium (column 5-6) P-wave states at T=0. Note that in the absence of velocity dependent corrections to the static potential, the $\chi_c(2P)$ state mass is not reproduced to the same accuracy than that of the other P-wave states.}\label{Tab:VacParm}
\end{table}

By adjusting $m_D$ it is possible to both reproduce the lattice values for $\Re[V]$ at finite temperature and to obtain at the same time a prediction for the values of $\Im[V]$, as shown by the solid lines in Fig.~\ref{Fig:ReImPot}. I.e. all information about the in-medium modification of $V(r)$ is contained in the value of $m_D$, which we list in Tab.~\ref{Tab:DebMass} and plot in Fig.~\ref{Fig:mDFits}. The red solid line shown in Fig.~\ref{Fig:mDFits} represents an interpolation function \cite{Arnold:1995bh,Vermaseren:1997fq} which connects to perturbative hard-thermal loop behavior at high temperatures, while allowing for non-perturbative deviations from the weak-coupling temperature dependence close to the phase transition.

To enable a realistic quarkonium phenomenology in the absence of a continuum limit extrapolation from the lattice, we proceed as follows. Using the renormalon subtracted \cite{Pineda:2001zq} bottom quark mass $m_b^{\rm RS'}=4.882\pm0.041 \,{\rm GeV}$, appropriate for a pNRQCD Schr\"odinger equation computation, we compute the energy levels of the corresponding Hamiltonian to reproduce the vacuum bottomonium S,P and D wave bound state masses. To be able to reproduce the experimentally observed bound state spectrum we amend Eq.\eqref{Eq:Cornell} by a string breaking term at $T=0$ that smoothly flattens off the linear rise at $\lambda_{\rm sb}=1.25$~fm. The best agreement to the PDG data is then reached with the following set of continuum $T=0$ parameters
\beq
c=-0.1767\pm 0.0210~{\rm GeV}, \quad \tilde\alpha_s=0.5043\pm 0.0298,\quad \sqrt{\sigma}=0.415\pm0.015~{\rm GeV}.\label{T0const}
\eeq
While the masses of the S-wave states are reproduced down to the third digit using the static potential alone, the reproduction of the P-wave states is less accurate as can be seen from the values summarized in Tab.~\ref{Tab:VacParm}. The absence of velocity and spin dependent corrections to the potential in our study is the reason for the worse quantitative agreement, besides the even more clear qualitative discrepancy in that the scalar, vector and tensor states remain degenerate in this calculation. For the charmonium sector we use the same $T=0$ parameters as those obtained for bottomonium and adjust the charm mass as fit parameter to reproduce the PDG S- and P-wave states, since the ensuing value of $m_c^{\rm PDG\,fit}=1.472\, {\rm GeV}$ cannot be reliably determined perturbatively in the renormalon subtraction scheme \cite{Pineda:2001zq}.

Combining the phenomenologically determined $T=0$ parameters with the continuum corrected Debye mass from the lattice $m_D^{\rm phys}(T)=m_D^{\rm lat}(\frac{T}{T^{\rm lat}_C} T^{\rm phys}_C) \frac{\sqrt{\sigma^{\rm phys}}}{ \sqrt{\sigma^{\rm lat}} }$ we obtain the finite temperature values of $\Re[V]$ and $\Im[V]$ from the extended Gauss Law ansatz, which in turn we use to solve the Schr\"odinger equation for the spectral function as detailed in Ref.~\cite{Burnier:2015tda}.

\begin{figure}[t!]
\centering
 \includegraphics[scale=0.35, trim= 0 2cm 0 1cm, clip=true]{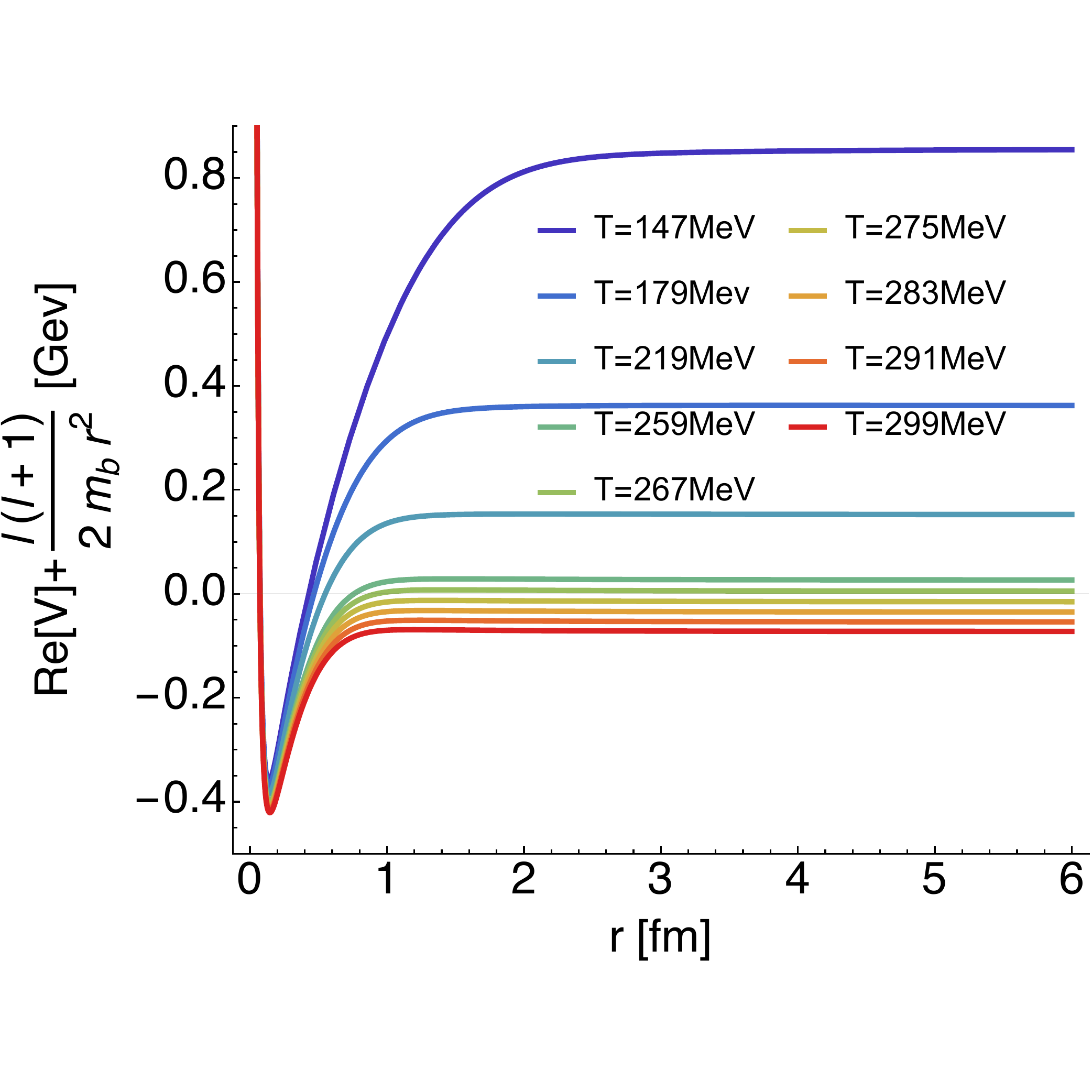}\hspace{0.2cm}
 \includegraphics[scale=0.35, trim= 0 2cm 0 1cm, clip=true]{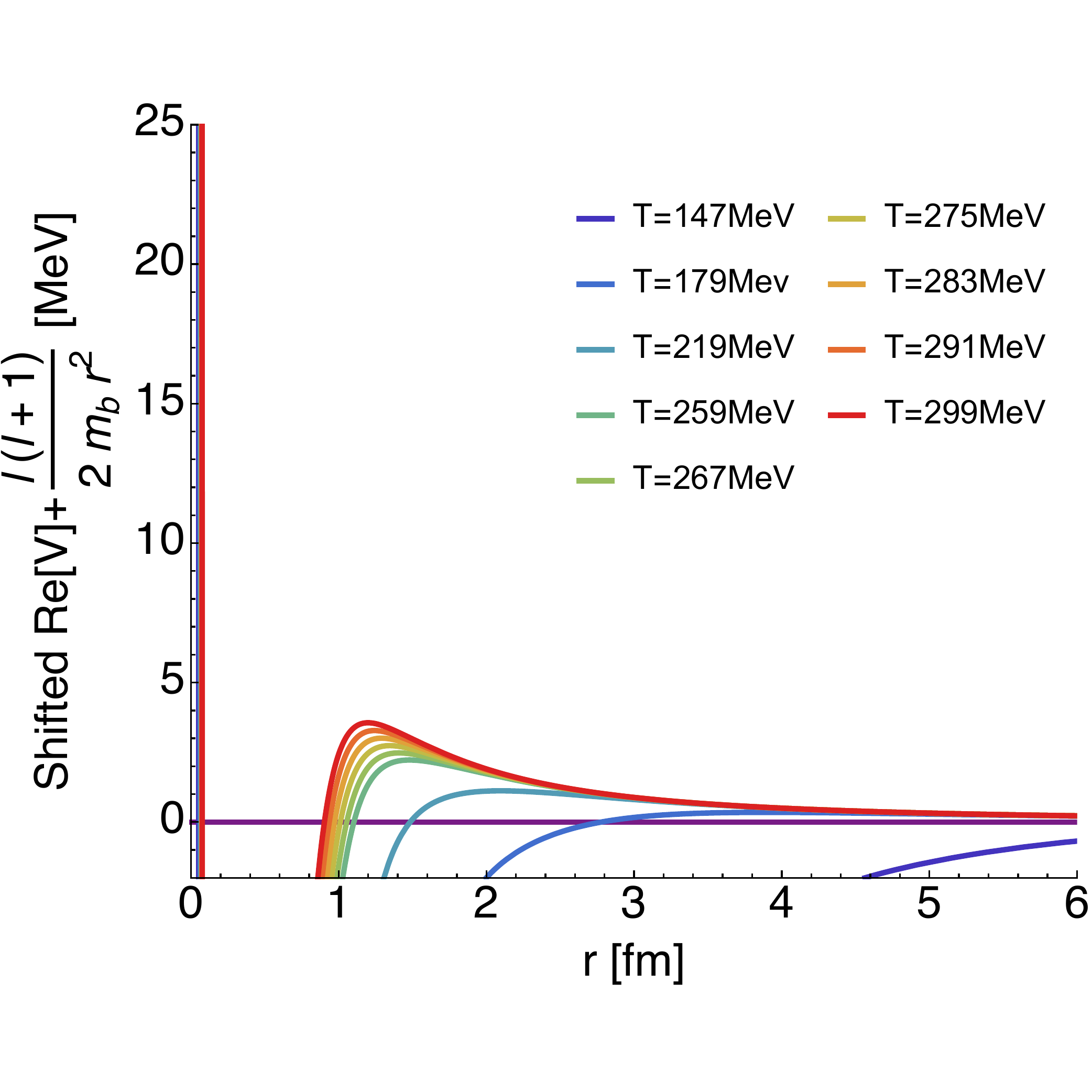}
 \includegraphics[scale=0.35, trim= 0 2cm 0 1cm, clip=true]{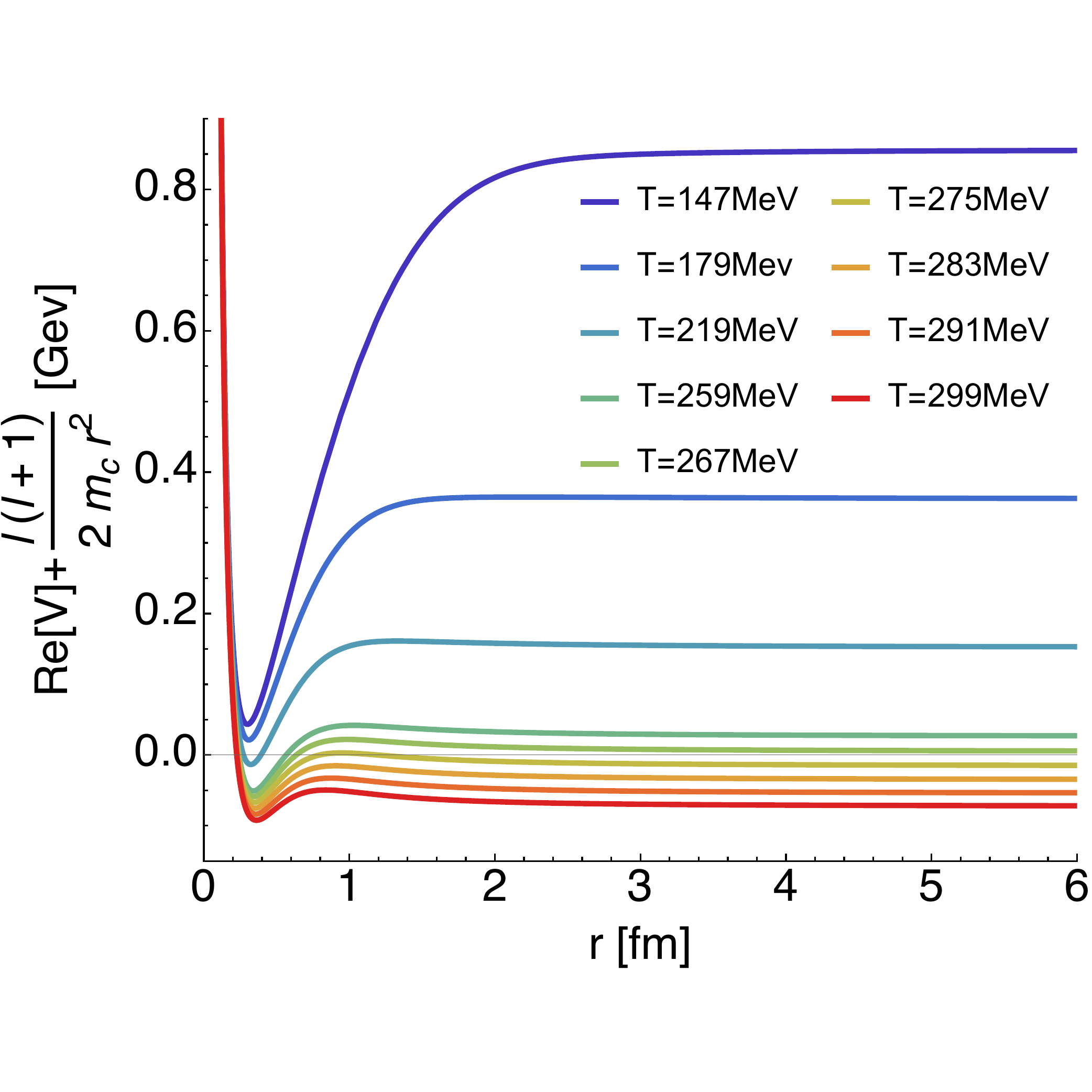}\hspace{0.2cm}
 \includegraphics[scale=0.35, trim= 0 2cm 0 1cm, clip=true]{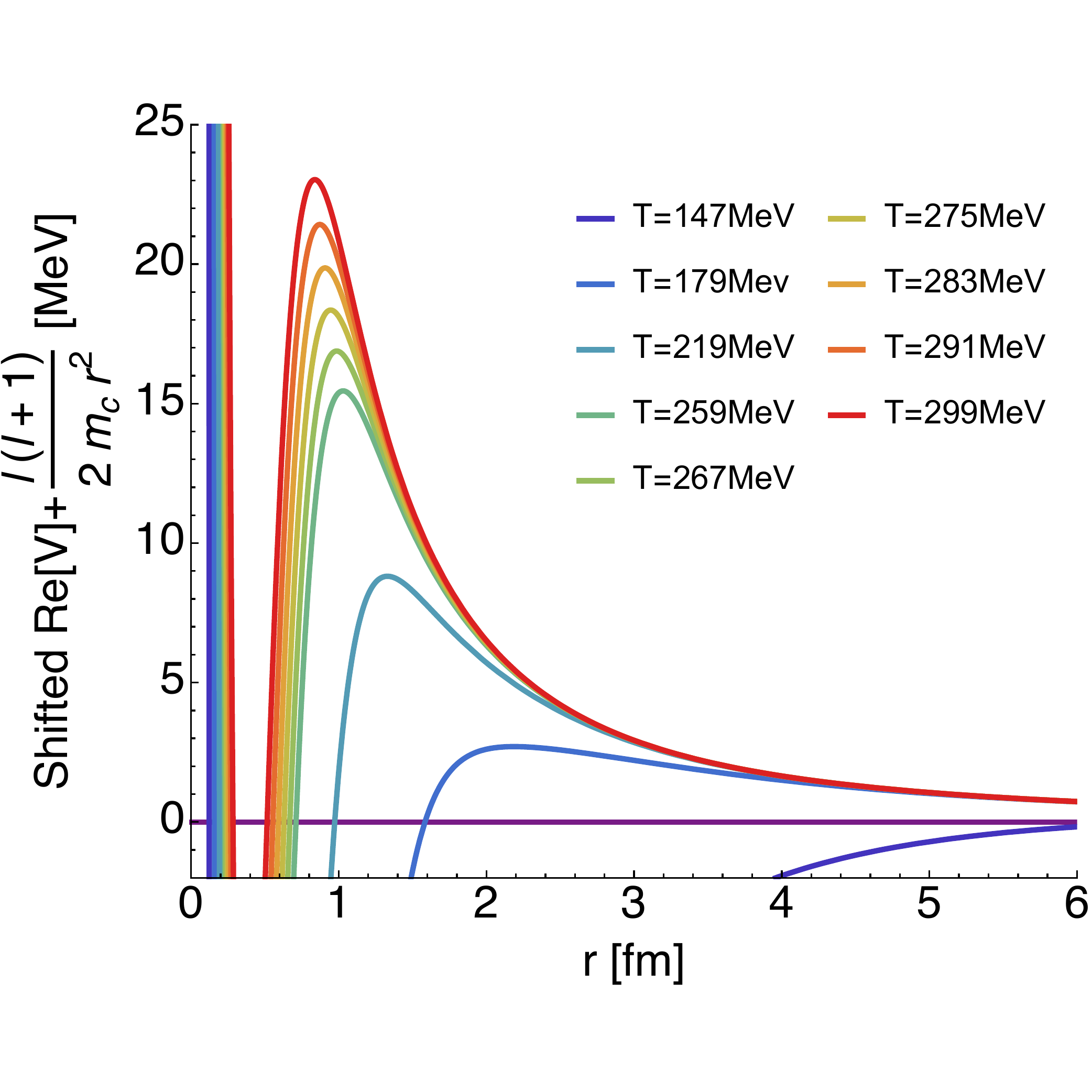}
 \caption{(left) The centrifugal term of the radial Schr\"odinger equation added to the real part of the in-medium potential at several temperatures around the transition temperature for bottomonium (top) and charmonium (bottom). (right) The sum of the two terms shifted by its value at infinite distance. Note the buildup of a centrifugal barrier, which does contribute significantly at $T=0$ and which is more pronounced for charmonium than for bottomonium due to the smaller constituent quark mass.}\label{Fig:CentrBarr}
\end{figure}

In the presence of finite angular momentum the reduced radial Schr\"odinger equation features a centrifugal term. As it enters with positive sign and contains an inverse quadratic dependence on the spatial distance it eventually compensates the Coulombic behavior of $\Re[V]$ and leads to a dip-like behavior at small distances (see left column of Fig.~\ref{Fig:CentrBarr}). Contrary to the case of molecular binding with e.g. van-der-Waals forces, at $T=0$ the combination of confining potential and centrifugal term does not lead to a significant centrifugal barrier. Interestingly however, once thermal fluctuations begin to weaken the linearly rising behavior of the real-part of the in-medium potential  and in particular when moving into the deconfinement transition, we find that a centrifugal barrier emerges dynamically (see right column of Fig.~\ref{Fig:CentrBarr}). As the finite angular momentum term is also weighted by the inverse quark mass its influence in absolute terms is larger for charmonium than for bottomonium, while its effect, i.e. the height of the centrifugal barrier is only a few MeV, for bottomonium $E^{\rm thresh}_b=4$~MeV at $T=300$~MeV and for charmonium $E^{\rm thresh}_c=23$~MeV. This additional feature of the in-medium potential in the presence of finite angular momentum will have a stabilizing effect on the bound states and we will find in the following section that it contributes to a characteristic difference in the survival patterns of resonance peaks within the in-medium spectra.

\subsection*{In-medium P-wave properties}

\begin{figure}
\centering
 \includegraphics[scale=0.365]{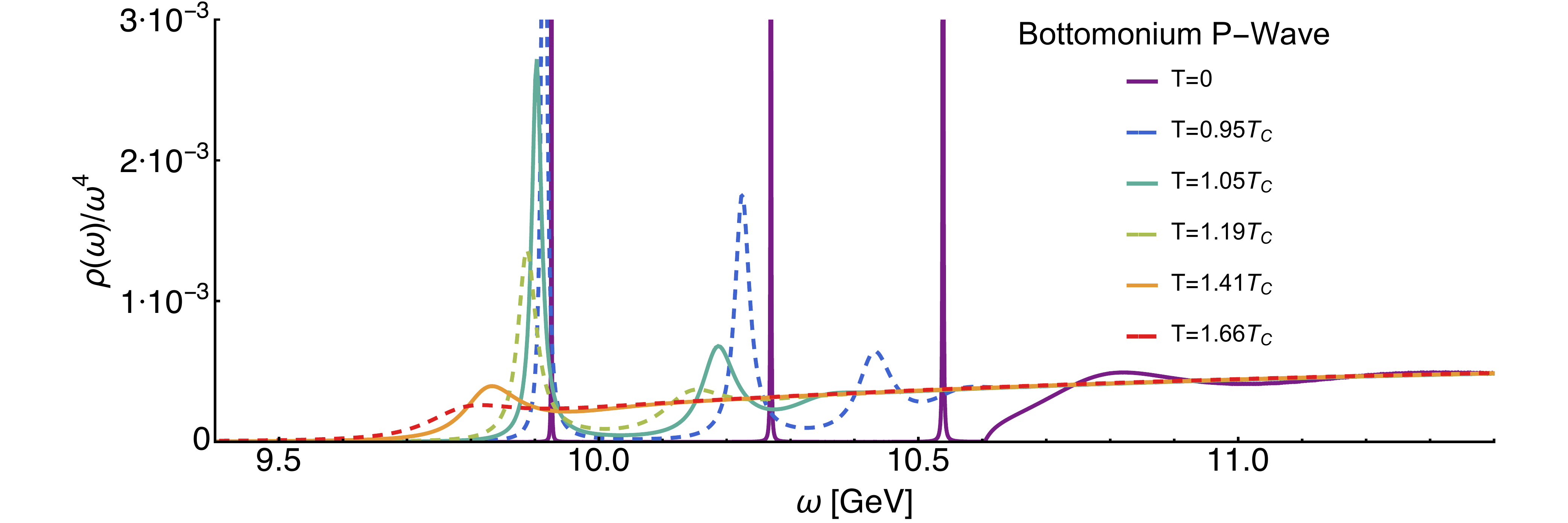}\hspace{0.2cm}\vspace{0.2cm}
 
 \includegraphics[scale=0.35]{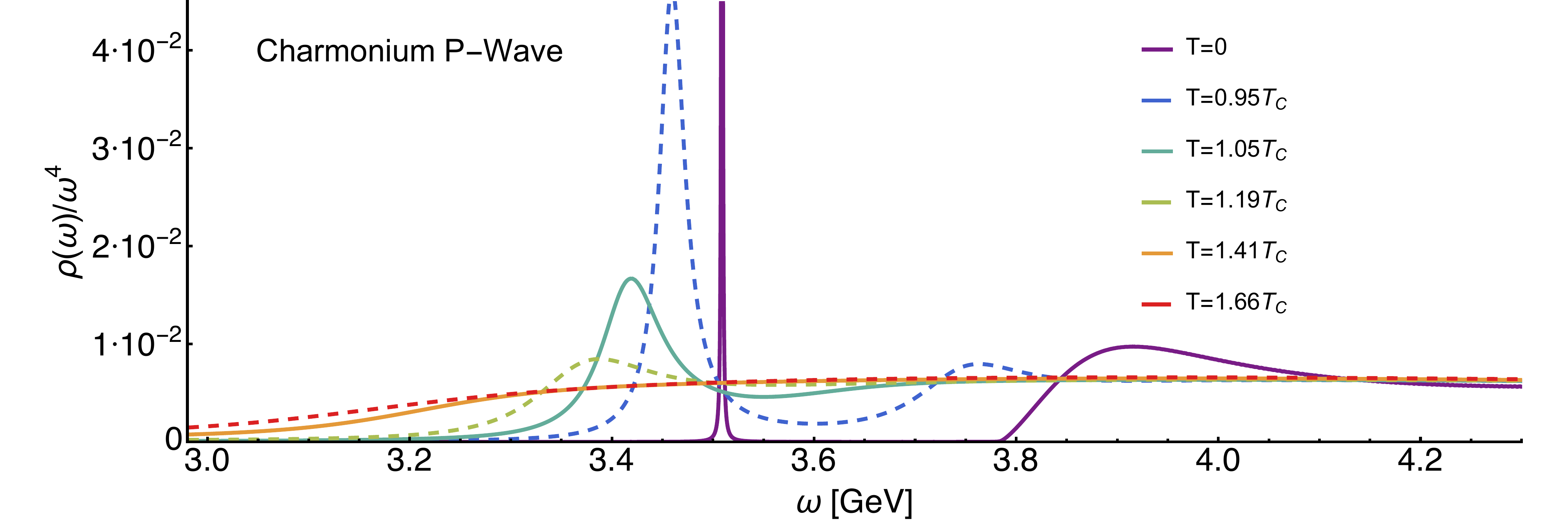}
 \caption{(top panel) In-medium bottomonium P-wave spectral function from our in medium potential based approach. Similar to the S-wave case we find clear indications for sequential melting with characteristic shifts of the peak position to lower values. While $\chi_b(3P)$ can still be identified as individual structure at $0.94T_C$ it has fully disappeared shortly above $T_C$. The $\chi_b(2P)$ peak reduces in strength by a factor ten between $1.05 T_C$ and $1.19T_C$.  For $T=1.41T_C$ only $\chi_b(1P)$ remains as distinguishable feature, which at our highest lattice temperature of $1.66T_C$ appears almost washed out. (bottom panel) The charmonium P-wave spectral functions based on the same complex potential as for bottomonium. As the $\chi_c(2P)$ state already in vacuum lies close to the continuum threshold its presence as threshold enhancement persists only up to $0.95T_C$. $\chi_c(1P)$ on the other hand induces a peaked structure up to $1.19T_C$ before disappearing completely.
 }\label{Fig:ChrmBotRhoOvervwP}
\end{figure}

Starting from the static potential, the quarkonium spectrum can be obtained using the formalism of Ref.~\cite{Burnier:2007qm} \footnote{It was found in \cite{Burnier:2007qm} that if the potential contains a Coulombic part, the S-wave might mix by a numerically small amount with the P-wave spectrum. Here we do not consider such a contribution but it may easily be introduced using the S-wave results of Ref.~\cite{Burnier:2015tda}.}.
In Fig.~\ref{Fig:ChrmBotRhoOvervwP} we present two summary plots for the P-wave spectral functions computed according to the lattice QCD potential based approach detailed above. Results for both bottomonium as well as charmonium at several temperatures close to the transition temperature are shown. We would like to remind the reader that the width of the peaks correspond to the physical value of the thermal width arising from medium fluctuations and that the bound state features collapses to delta-peak structures at vanishing temperature and are not numerical artefacts\footnote{In this non-relativistic pNRQCD scenario, neither electromagnetic, nor Hadronic decays are included, so that at T=0 a true delta peak like behavior ensues. We have therefore added a small but finite imaginary part at $T=0$ so that the vacuum peaks in Fig.~\ref{Fig:ChrmBotRhoOvervwP} can be visualized.}. The positions of the peaks at $T=0$ correspond to the masses listed in Tab.~\ref{Tab:VacParm}.

An inspection by eye already reveals both qualitative similarities as well as differences between the P-wave and the S-wave in-medium spectra. While the more deeply bound bottomonium states are affected less strongly compared to charmonium, all states show a characteristic broadening and shifts of the peak position to lower values with increasing temperature. One major difference is the behavior of the continuum. Just as in the absence of finite angular momentum the threshold starts to close in on the spectral peaks sequentially from above, for the P-wave however it passes through the bound state features before they are washed out and become part of that continuum.

\begin{figure}
\centering
 \includegraphics[scale=0.35, trim= 0 1.5cm 0 1cm, clip=true]{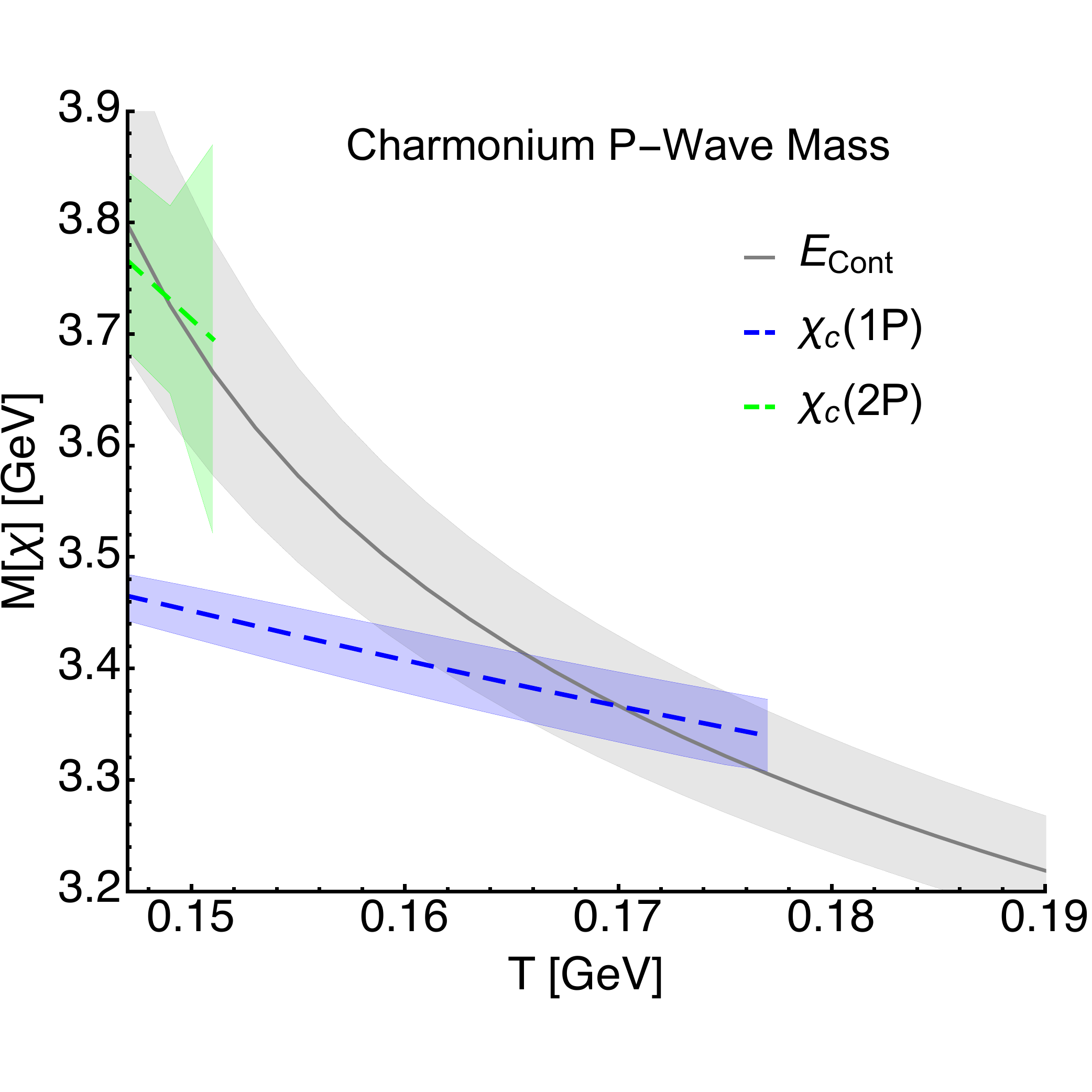}\hspace{0.2cm}
 \includegraphics[scale=0.35, trim= 0 1.5cm 0 1cm, clip=true]{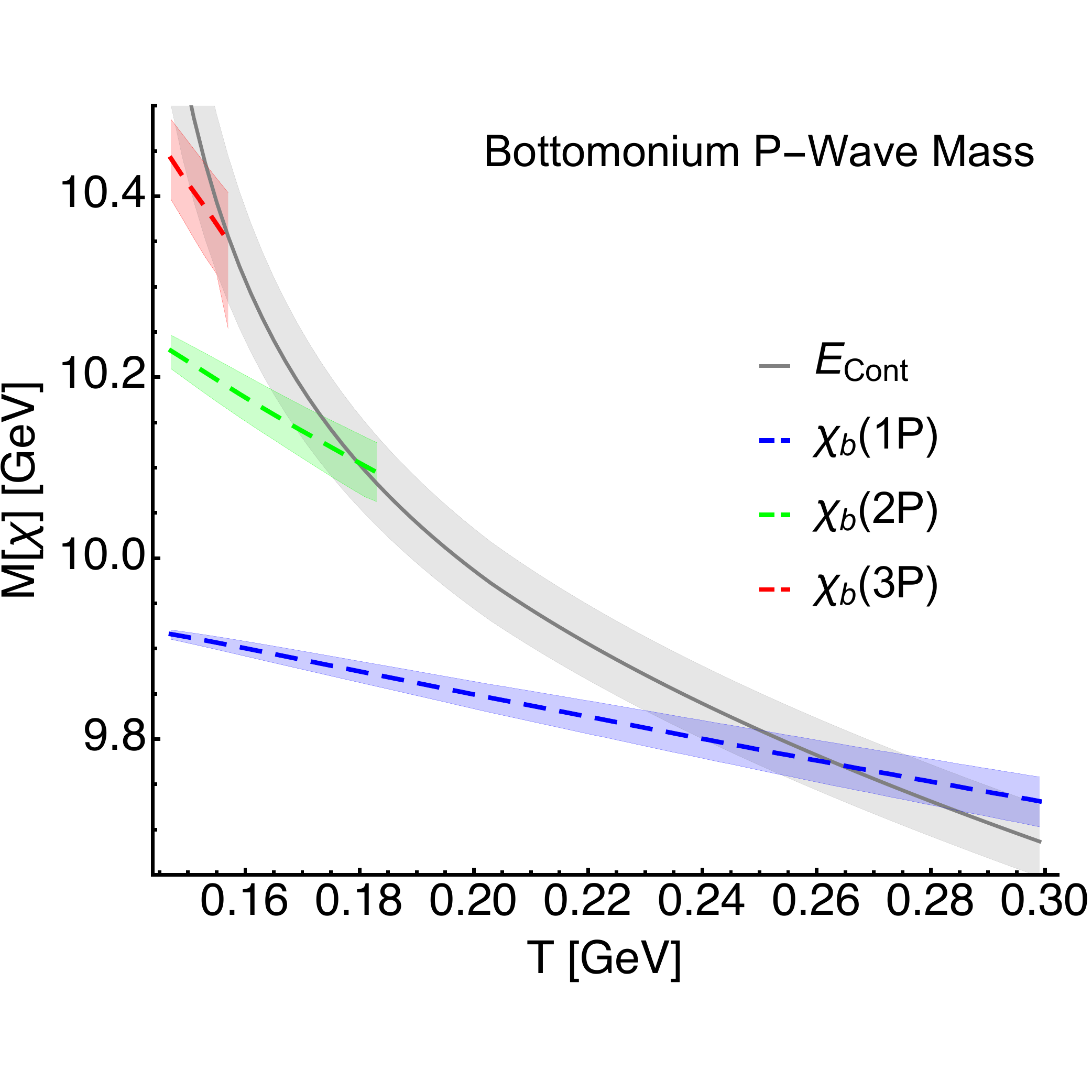}
 \caption{Mass of the P-wave charmonium (left) and bottomonium (right) bound states from the position of their in-medium spectral peaks. As in the S-wave case we find that their peak position decreases monotonously with temperature until the bound state disappears. Different from the case of vanishing angular momentum however, peaks can be identified even in those cases where the continuum threshold has already moved below their position. I.e. the colored curves can cross the gray continuum line. For the S-wave such crossing was avoided before disappearance of the peak. The error bands reflect the uncertainty  stemming from the Debye mass determination.
}\label{ET}
\end{figure}

\begin{figure}
\centering
 \includegraphics[scale=0.35, trim= 0 1.5cm 0 1cm, clip=true]{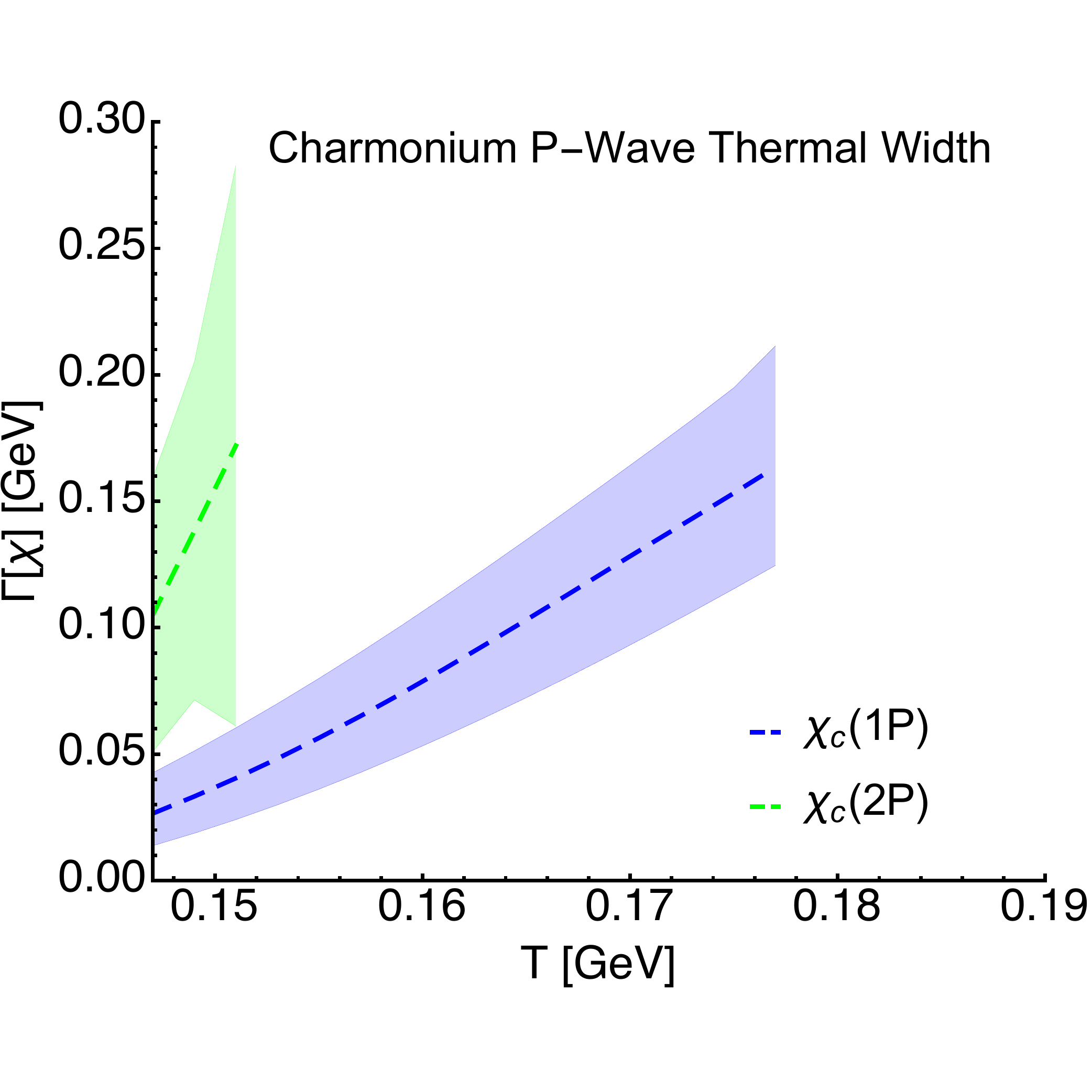}\hspace{0.2cm}
 \includegraphics[scale=0.35, trim= 0 1.5cm 0 1cm, clip=true]{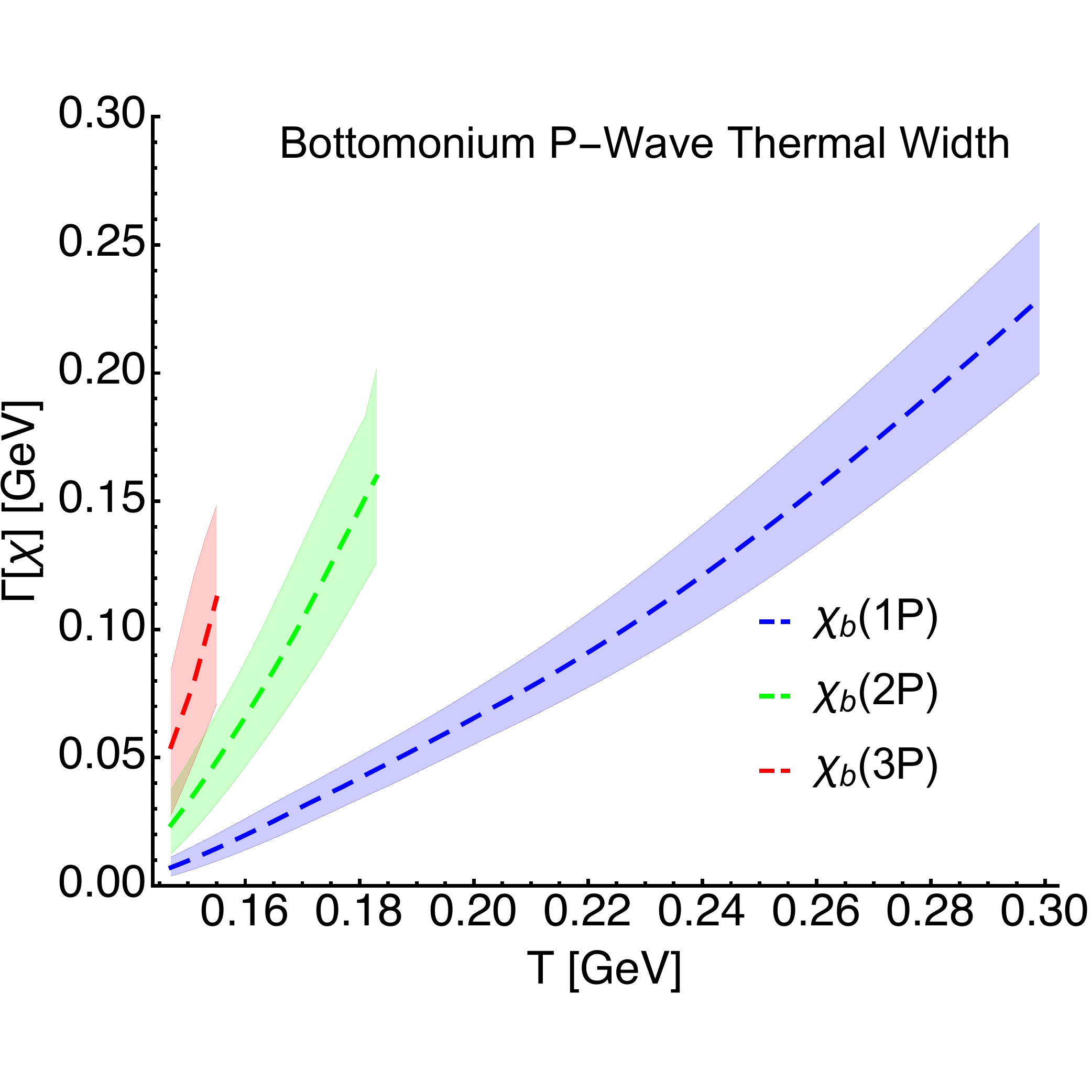}
 \caption{Width of the P-wave charmonium (left) and bottomonium (right) bound states, which increases monotonously with temperature. The error bands reflect the uncertainty stemming from the Debye mass determination.
}.
\label{WidthT}
\end{figure}

\begin{figure}
\centering
 \includegraphics[scale=0.35, trim= 0 1.5cm 0 1cm, clip=true]{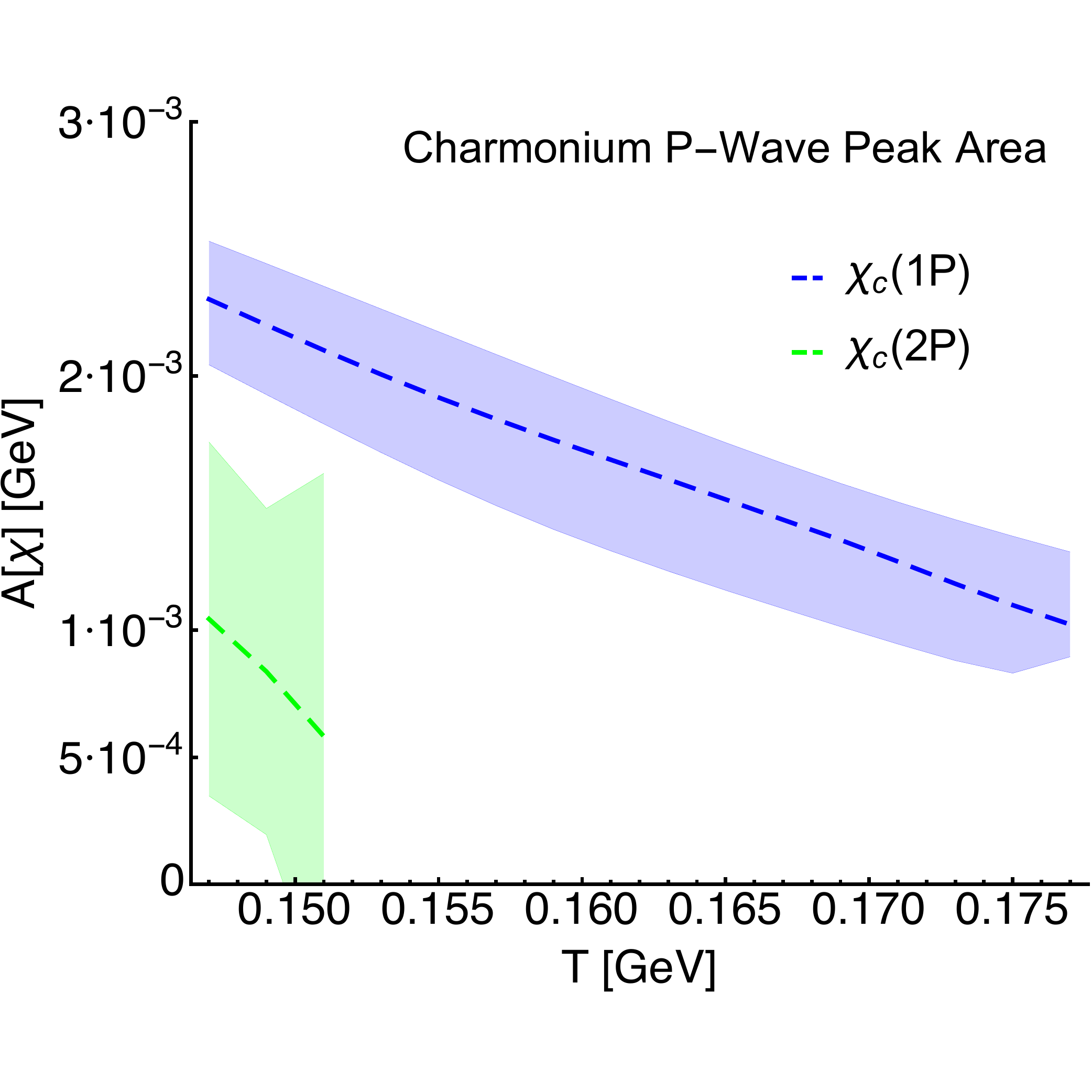}\hspace{0.2cm}
 \includegraphics[scale=0.35, trim= 0 1.5cm 0 1cm, clip=true]{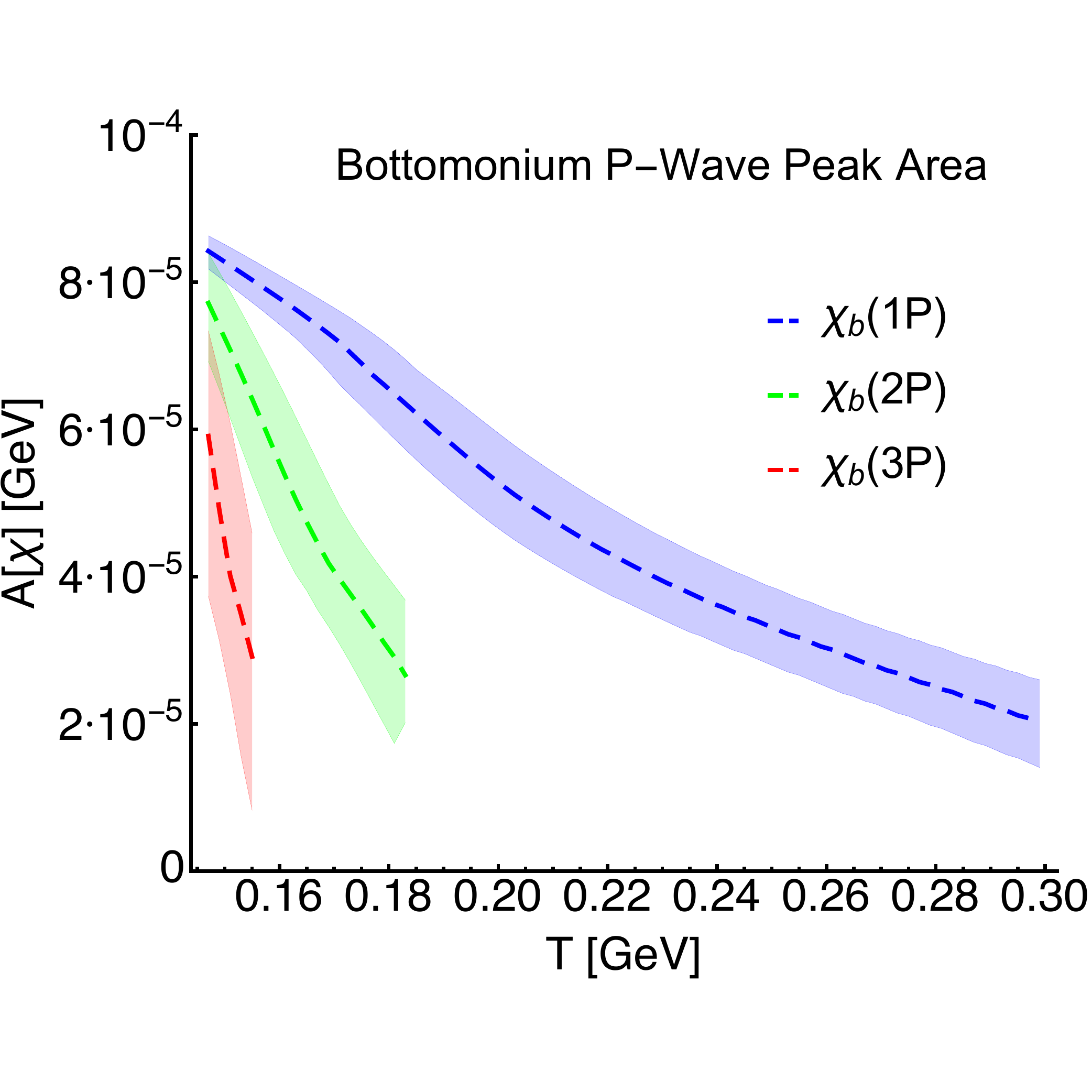}
 \caption{Area under the bound-state peaks in the P-wave charmonium (left) and bottomonium (right) spectrum. The plateau structure seen in the S-wave case is much less pronounced here. Error bands reflect the uncertainty stemming from the Debye mass determination.}\label{A}
\end{figure}

For a more quantitative assessment of this behavior let us carry out a finely spaced $\delta_T=2$~MeV scan of the available temperature regime and fit the spectral features using a skewed Breit-Wigner as suggested \cite{Taylor} by scattering theory 
\beq
\rho(\omega\approx E)=C \frac{(\Gamma/2)^2}{(\Gamma/2)^2+(\omega-E)^2}+2
\delta \frac{(\omega-E)\Gamma/2}{(\Gamma/2)^2
+(\omega-E)^2} +\sigma_1+\sigma_2(\omega-E)+\mathcal{O}(\delta^2),\label{sBW}
\eeq
where $E$ denotes the energy of the in-medium excitation, $\Gamma$ its width and
$\delta$ the phase shift. In addition to these parameters we include $\sigma_1$ and $\sigma_2$ to account for the presence of possible threshold artefacts. We are interested in the temperature dependence of the peak position, width and integrated area, which are plotted in Fig.~\ref{ET}, \ref{WidthT} and \ref{A}. Note that that area in particular is of phenomenological relevance as it is related to the in-medium decay of the state either into diphotons or light hadrons. 

Our initial observations are clearly resembled in the quantitative plots. The masses of the individual states in Fig.~\ref{ET} show a monotonous decrease with temperature, however in contrast to the S-wave case we find that e.g. the ground state peak feature can be clearly identified at temperatures, where the continuum has already moved below its position. Such a bound state remnant in-medium excitation should be considered as a metastable resonance, which cannot be attributed a positive binding energy in the common sense. At the same time the thermal width in Fig.~\ref{WidthT} shows a strong monotonous increase.

With the interplay between state broadening and being engulfed by the continuum, the integrated area of individual peaked features is more difficult to determine in the P-wave than in the S-wave. Fig.~\ref{A} furthermore shows that for the finite angular momentum states we do not find a similarly pronounced plateau for the excited states and even in the ground states the area seems to be on a continuous decline with increasing temperature. 

\subsection*{Melting temperatures}

The preceding sections have shown that the question of what constitutes quarkonium melting is a subtle question for the P-wave states. On the one hand we have the fact that spectral features seem to persist at temperatures where they are embedded into the continuum. This tells us that remnants of these former bound states will still lead to a deviation from naive thermal quarkonium production predictions, even if they are classified as melted according to their conventional binding energy.  On the other hand the peak area did not show a behavior akin to a plateau, neither for the $\chi_b(1P)$ ground state nor for $\chi_c(1P)$ or all other excited states, where instead a monotonous decrease is observed. To define melting in this case, we have to specify a quantitative value for the area at which the influence of the state is to be disregarded, in the following we choose $A=\frac{1}{2}A(T=0)$.

In the absence of an unambiguous criterion for melting, we list in Tab.~\ref{t::melting} the dissolution temperatures both according to the conventional condition when the in-medium binding energy $E_{\rm bind}$ of the state equals its thermal width and the area condition introduced above. $E_{\rm bind}$ is defined from the difference between the in-medium mass of the state and the asymptotic value of the real-part of the potential at large distances, which does not take into account the effect of the centrifugal barrier. I.e. the values from the binding energy in Tab.~\ref{t::melting} provide lower limits to the melting temperature.

\begin{table}[h!]
\centering
 \begin{tabular}{|c|c|c||c|c|c|}\hline
states& $ \chi_c(1P)$& $\chi_c(2P)$ & $\chi_b (1P) $ &$\chi_b (2P) $ &$\chi_b (3P) $ \\ \hline\hline
 $T_{\rm melt}^{\Gamma=E_{\rm bind}}/T_C $ &$1.04(3) $& $<0.95$  & $1.41(6)$ & $1.06(3)$ & $0.98(2)$ \\   \hline
 $T_{\rm melt}^{A=\frac{1}{2}A_0}/T_C $ &$1.18(8)$& $<0.95$  & $1.72(13)$ & $1.08(5)$ & $0.98(3)$ \\   \hline
 \end{tabular}
 \caption{Melting temperatures $T_{\rm melt}$ of the different bound states defined (top row) by the point, at which the width of the state equates its binding energy and (bottom row) where the area of the in-medium peak becomes half of its size at $T=0$. The error on the determination of $T_{\rm melt}$ takes into account the possible variation of $m_D$ as shown in Fig.~\ref{Fig:mDFits}.}\label{t::melting}
\end{table}

It is clear from Tab.~\ref{t::melting} that for the excited states the two definitions lead to very similar values of the melting temperature. For the ground state however the results differ in a statistically significant way, with the melting temperature according to the area condition being larger than the conventionally defined $T_{\rm melt}^{\Gamma=E_{\rm bind}}$. Since the latter does not take into account the effects of the centrifugal barrier, arising at finite angular momentum, i.e the binding energy is defined with respect to $\Re[V]$ at infinite distance, it should indeed be considered rather as a lower bound to the melting temperature for the P-wave. The area on the other hand does reflect the stabilizing effect of the finite angular momentum, which makes the corresponding definition of the melting temperature more appropriate for the P-wave.

Our finding should be compared to the results from two recent investigations of in-medium quarkonium spectral functions using the effective theory of NRQCD on the lattice \cite{Aarts:2014cda,Aarts:2012ka,Skullerud:2014sla,Kim:2014iga,Kim:2014nda,Kim:2015csj,Kim:2015rdi}. The benefit of this effective field theory approach lies in the inclusion of finite velocity and spin dependent contributions that do not enter the spectral function computations based on a static potential. The drawback on the other hand is that in lattice NRQCD, quarkonium spectra need to be reconstructed from Euclidean simulation data directly via Bayesian inference. The quality of available lattice NRQCD data, i.e. the number of datapoints $N_\tau$ and precision $\Delta D/D$, as well as the fact that the underlying problem is inherently ill-posed have only allowed the study of the S-wave and P-wave ground states for bottomonium and charmonium in that approach. Spectral peak positions are reliably captured, but a dependable determination of spectral widths is very challenging with a Bayesian spectral reconstruction and has so far not been achieved in a quantitatively robust manner.  In the absence of such a quantitative evaluation of the thermal width, melting of a state in these studies has so far been defined by the disappearance of peaked structures in the reconstructed spectra. Our findings especially for the P-wave show that such a definition might lead to ambiguous outcomes, since peaked structures may survive up to much higher temperatures than the melting point as defined from the comparison between binding energy and width. 

And indeed, while both NRQCD studies agreed that the peak feature of the S-wave bottomonium ground state $\Upsilon(1S)$ survives deep into the QGP phase up to at least $T=1.9T_C$, their conclusions for the P-wave differ. Work based on the Maximum Entropy Method suggested a disappearance of the $\chi_b(1P)$ peak shortly above the deconfinement crossover transition, while an investigation using a novel Bayesian spectral reconstruction method hinted at the survival of a remnant bound state structure up to $T=1.56T_C$. Compared to the outcome of the present work, as shown in Fig.~\ref{Fig:ChrmBotRhoOvervwP} and \ref{ET}, the MEM results seems to underrepresent peak structures that in our results is still clearly visible even at $T=1.41T_C$. On the other hand at $T=1.56T_C$, where the novel Bayesian method identified a remnant peak, it is not unfathomable from Fig.~\ref{Fig:ChrmBotRhoOvervwP} that a weak but distinguishable feature remains in the spectrum. Of course the inclusion of velocity and spin dependent corrections to the static potential may either weaken or strengthen the remnant bound state features, therefore is is paramount to pursue an evaluation of such higher order contributions to the potential in the future.

Note that when comparing charmonium and bottomonium using the same $T_{\rm melt}^{\Gamma=E_{\rm bind}}$ we find that the value for $\chi_b(1P)$ lies close to the one for $T_{\rm melt}^{J/\psi}/T_C=1.37\pm0.08$ for $J/\Psi$, which is in in accord with expectations, since the vacuum binding energies of the two states are nearly identical, with the P-wave being slightly more weakly bound. Recent studies of charmonium correlation functions in lattice NRQCD also showed in-medium modification hierarchically ordered according to the vacuum binding energy \cite{Kim:2015csj,Kim:2015rdi}. 

\section{Implications for heavy-ion collisions}
\label{Sec:HeavyIon}

The intricate suppression patterns observed for S-wave bottomonium \cite{Chatrchyan:2011pe,Chatrchyan:2012lxa,Moon:2014lia} and charmonium \cite{Adare:2006ns,Adare:2011yf,Adamczyk:2012ey,Abelev:2013ila,Adam:2015isa,Adam:2015gba} states at RHIC and LHC have significantly advanced our understanding of the in-medium modification of quarkonium states in relativistic heavy-ion collisions. The contribution of P-wave feed down to the final S-wave abundances however remains a piece of the overall puzzle. Here we attempt to estimate the fraction of feed-down from in-medium $\chi_b$ and $\chi_c$ states to the corresponding S-wave states in a heavy-ion collision. To this end we assume that the heavy quarks are fully kinetically thermalized at the phase boundary. For charm quarks this reasoning e.g. underlies the successful statistical model of hadronization \cite{Andronic:2009sv}, as well as estimates for the kinetic equilibration of heavy quarks  \cite{Francis:2015daa} and is supported by the measurement of a finite elliptic flow for $J/\psi$ mesons \cite{ALICE:2013xna}. In the heavier bottomonium sector it is less clear how well a kinetic equilibration picture is applicable, in particular, since up to now observables that could indicate collective behavior have not been measured in detail.

With this caveat in mind let us proceed. The quantity of phenomenological interest is the fraction of S-wave vacuum states that arise from the radiative decay of P-wave states, which in $p+p$ collisions is defined \cite{Aaij:2014caa} as 
\begin{align}
_{pp}{\cal R}^{\chi(mP)}_{\psi (nS)}\equiv \frac{\sigma(pp\to\chi(mP)X)}{\sigma(pp\to\psi(nS)X)} {\cal{B}}_{\chi\to\psi\gamma} = \frac{N(pp\to\chi(mP)X)}{N(pp\to\psi(nS)X)} {\cal{B}}_{\chi\to\psi\gamma},
\end{align}
where $\chi$ either stands for $\chi_b$ or $\chi_c$ and $\psi$ denotes the $\Upsilon$ states or $J/\psi$ respectively. $\cal{B}$ is the vacuum branching fraction of the radiative decay that embodies the feed-down contribution to the corresponding S-wave state. The value of $_{pp}{\cal R}^{\chi(mP)}_{\psi (nS)}$ have been recently measured in detail at $\sqrt{s}=7$TeV and $\sqrt{s}=8$TeV by the LHCb collaboration \cite{LHCb:2012af,Aaij:2014caa}. It was found that a significant portion, i.e. up to around 30\% of the S-wave states may actually originate in states with finite angular momentum. 

To obtain the corresponding values for a heavy-ion collisions, we will compute an approximation to the number ratio 
\begin{align}
\kappa^{\chi(mP)}_{\psi(nS)} =  \frac{N(AA\to\chi)}{N(pp\to\chi)} / \frac{N(AA\to\psi)}{N(pp\to\psi)} = \frac{N(AA\to\chi)}{N(AA\to\psi)} \frac{N(pp\to\psi)}{N(pp\to\chi)}\label{Eq:Korr}
\end{align}
at freezeout. It in turn provides us access to
\begin{align}
_{AA}{\cal R}^{\chi(mP)}_{\psi(nS)}\equiv \frac{N(AA\to\chi(mP)X)}{N(AA\to\psi(nS)X)} {\cal{B}}_{\chi\to\psi\gamma} =\kappa^{\chi(mP)}_{\psi(nS)}  \cdot {_{pp}{\cal R}}^{\chi(mP)}_{\psi (nS)}.\label{Eq:KorrVacFrac}
\end{align}

Let us proceed to estimate the number ratio $\kappa$, for which we follow a similar path as in our investigation of the $\psi'$ to $J/\psi$ ratio \cite{Burnier:2015tda}. It amounts to assuming an instantaneous freezeout scenario, where at the crossover temperature $T=T_C$ the area under the appropriately normalized in-medium spectral peaks is related to the number of vacuum states this area correspond to. For the S-wave channel the spectral area at finite temperature and the rate of dilepton emission in the medium are intimately related \cite{McLerran:1984ay}. In vacuum this rate is given by the square of the wavefunction at the origin \cite{Bodwin:1994jh}

\beq
R_{\ell\bar\ell}^{\psi(nS)} (T=0) \propto \frac{|\Psi_{\psi(nS)}(0)|^2}{M_{\psi}^2}, \quad  R_{\ell\bar\ell}^{\psi(nS)}(T>0)\propto \int dp_0 d^3\bp \frac{\rho(P)}{P^2}n_B(p_0).
\eeq

In case of the P-wave the spectral area on the other hand can be related to the decay of a $Q\bar{Q}$ color singlet into light hadrons \cite{Bodwin:1994jh}. At $T=0$ this process is governed by the derivative of the wavefunction at the origin. I.e. at finite temperature one needs to multiply the probability of the decay process with the probability that the corresponding quarkonium state is occupied, leading to

\beq
R_{LH}^{\chi(mP)} (T=0) \propto \frac{|\Psi^\prime_{\chi(mP)}(0)|^2}{M_{\chi}^4}, \quad  R_{LH}^{\chi(mP)}(T>0)\propto \int dp_0 d^3\bp \frac{\rho(P)}{P^4}n_B(p_0).
\eeq

We then combine the above expressions to compute 
\begin{align}
\kappa^{\chi(mP)}_{\psi(nS)}  =  \frac{R_{LH}^{\chi(mP)} (T>0)}{R_{LH}^{\chi(mP)} (T=0)} / \frac{R_{\ell\bar\ell}^{\psi(nS)} (T>0)}{R_{\ell\bar\ell}^{\psi(nS)} (T=0)}.
\end{align}

Since at the pseudo-critical temperature we still find well defined peaks for the states under consideration here, the integral over the in-medium spectrum can be simplified.  We use the fact that the spectral function to leading order only depends on $P^2=p_0^2-\bp^2$, so that after a variable transformation $\omega=\sqrt{p_0^2-\bp^2}$ the in-medium peaks can be replaced by a delta function positioned at the in-medium mass of the corresponding state $M_n$ multiplied by the peak area $A$ 
\beq
R(T>0) \propto A  \int d^3\bp\, n_B(\sqrt{M_n^2+\bp^2})\frac{M_n}{\sqrt{M_n^2+\bp^2}}.
\eeq

Using the fitted values of the peak area depicted in Fig.~\ref{A} together with the in-medium mass of Fig.~\ref{ET} we find the correction factors listed in Tab.~\ref{t:feeddowncorrT}
\begin{table}[h!]
\centering
 \begin{tabular}{|c|c|c||c|c||c| || c|}\hline
$\kappa^{\chi_b(1P)}_{\Upsilon (1S)} $ & $\kappa^{\chi_b(2P)}_{\Upsilon (1S)}$  & $\kappa^{\chi_b(3P)}_{\Upsilon (1S)}$  & $\kappa^{\chi_b(2P)}_{\Upsilon (2S)}$  & $\kappa^{\chi_b(3P)}_{\Upsilon (2S)} $ & $\kappa^{\chi_b(3P)}_{\Upsilon (3S)} $ & $\kappa^{\chi_c(1P)}_{J/\psi (1S)} $ \\ \hline\hline
$0.540(5)$& $0.068(4)$  & $0.012(2)$ & $0.250(1)$ & $ 0.043(7)$& $ 0.031(5)$ & $0.120(25)$ \\   \hline
 \end{tabular} 
\caption{Estimates of in-medium correction factor to the $p+p$ feed-down fractions from purely thermal modification of the in-medium quarkonium spectra.}\label{t:feeddowncorrT}
\end{table}

\begin{figure}
 \includegraphics[scale=0.3, trim= 0 1.5cm 0 1cm, clip=true]{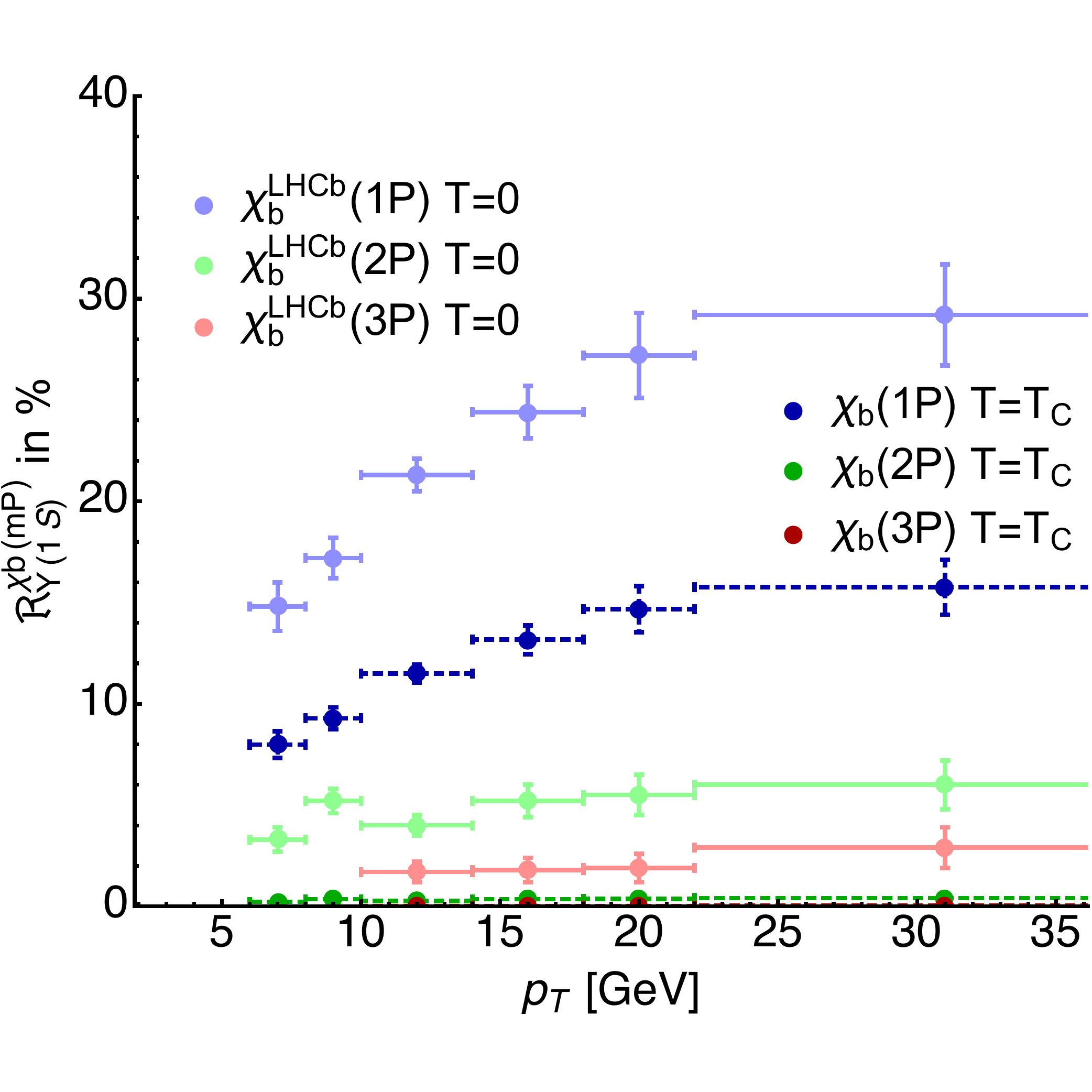}\hspace{0.2cm}
 \includegraphics[scale=0.3, trim= 0 1.5cm 0 1cm, clip=true]{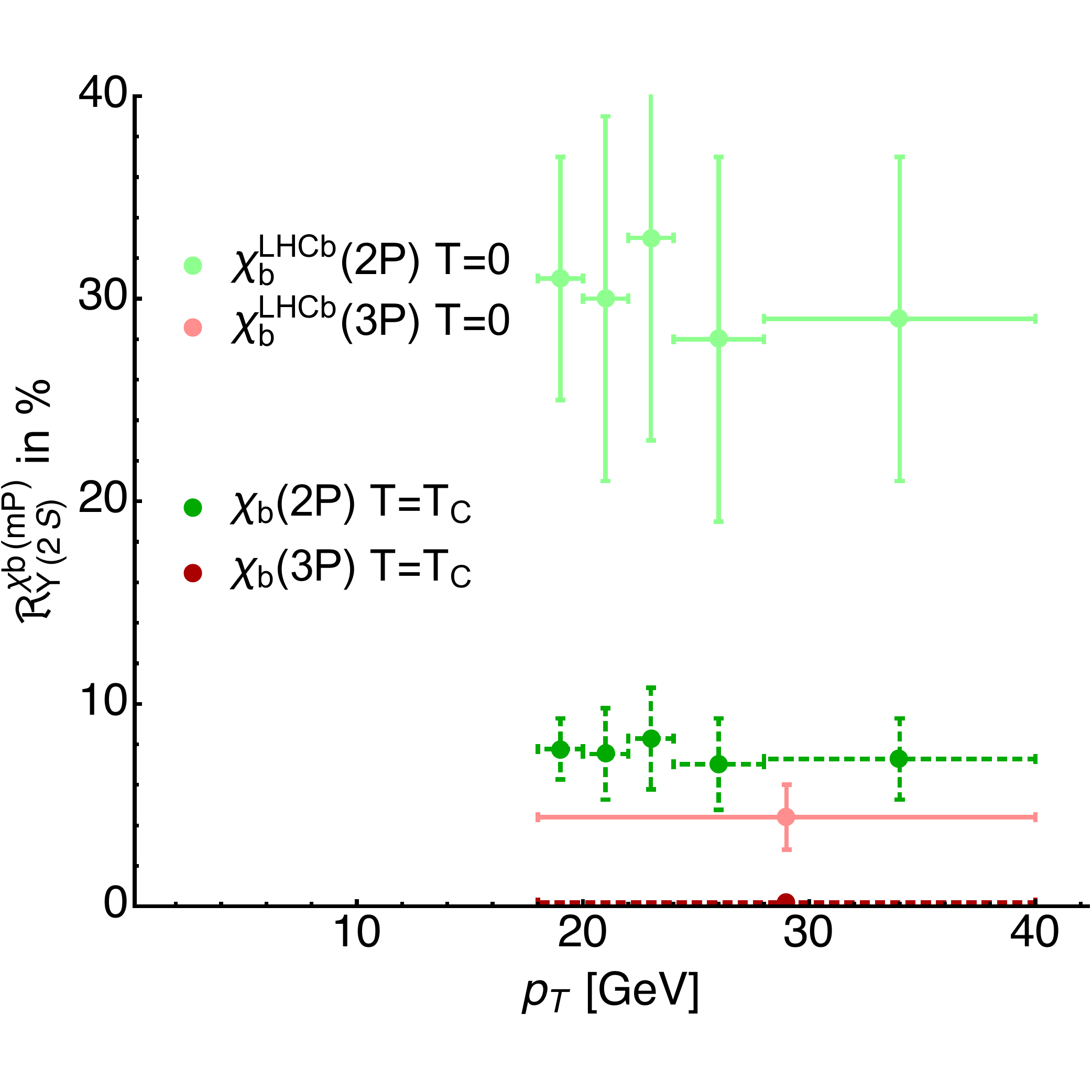}
 \includegraphics[scale=0.3, trim= 0 1.5cm 0 1cm, clip=true]{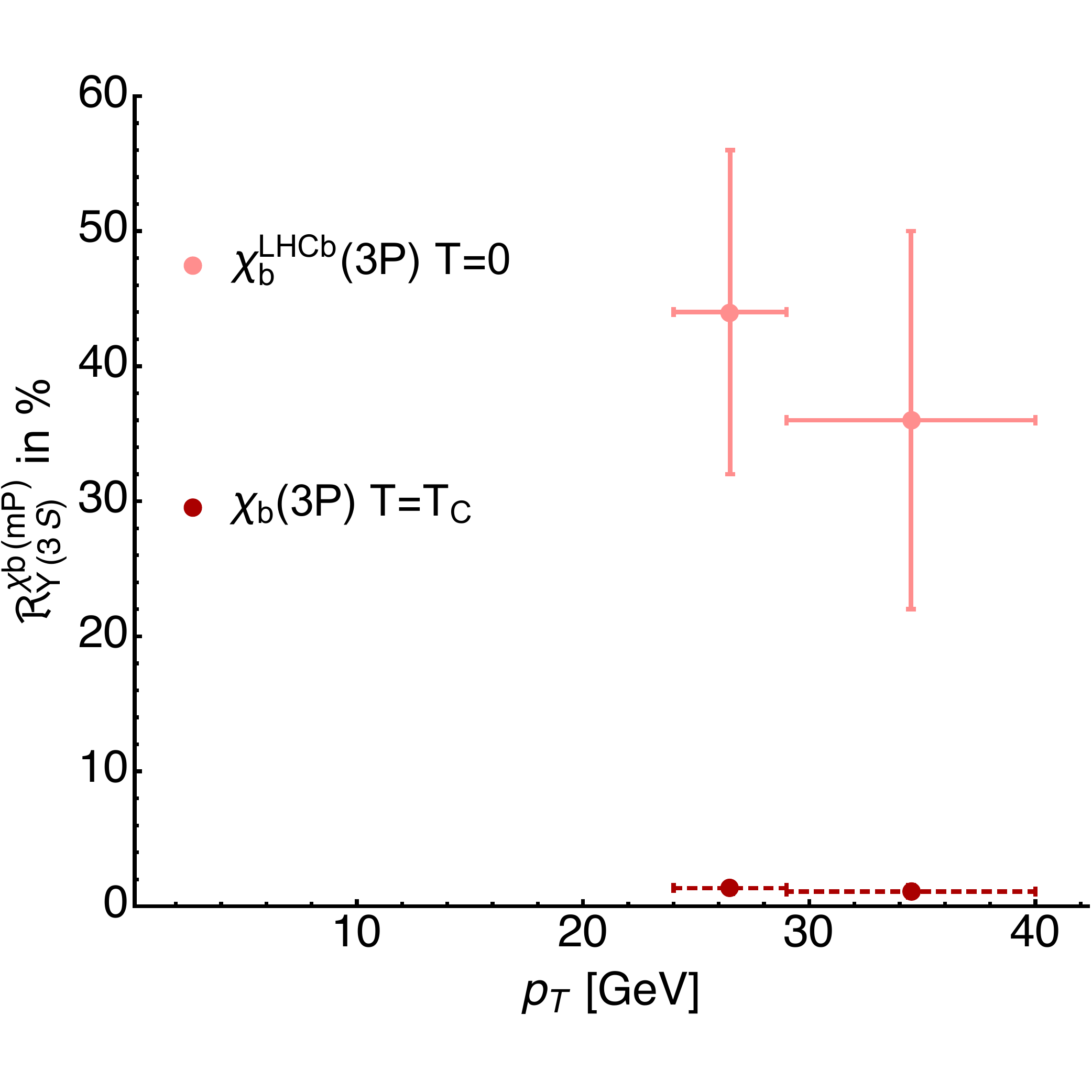}
 \caption{Comparison of our estimates for in-medium feed-down based on a purely thermal modification of the in-medium quarkonium spectra with the feed-down fractions for bottomonium in $p+p$ as measured by the LHCb collaboration at $\sqrt{s}=7$TeV \cite{Aaij:2014caa}.}\label{FeedDownUpsilon}
\end{figure}

\begin{figure}
\centering
 \includegraphics[scale=0.3, trim= 0 1.5cm 0 1cm, clip=true]{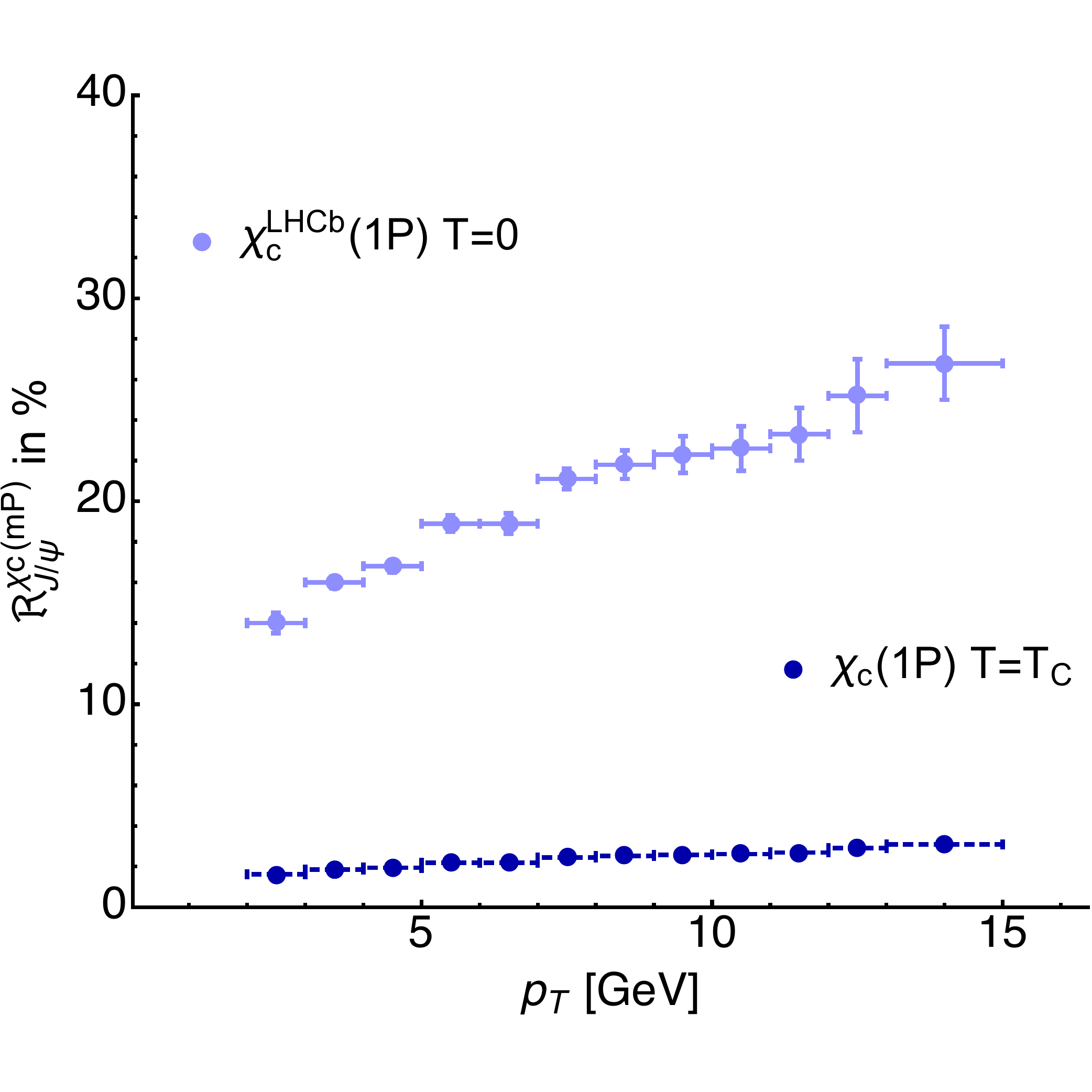}\hspace{0.2cm}
 \caption{Comparison of our estimates for in-medium feed-down based on a purely thermal modification of the in-medium quarkonium spectra with the feed-down fractions for charmonium in $p+p$ as measured by the LHCb collaboration at $\sqrt{s}=7$TeV \cite{LHCb:2012af}.}\label{FeedDownCharm}
\end{figure}

In order to compute an estimate for the in-medium feed-down fractions for heavy-ion collisions $_{AA}{\cal R}^{\chi(mP)}_{\psi(nS)}$, we obtained the $p+p$ data at $\sqrt{s}=7$TeV on $_{pp}{\cal R}^{\chi(mP)}_{\psi (nS)}$ by the LHCb collaboration from the hepData archive (see link for references \cite{LHCb:2012af,Aaij:2014caa}).  Inserting their measurements into Eq.\eqref{Eq:KorrVacFrac}, using the same correction factor at each $p_T$, we obtain the results plotted in Fig.~ \ref{FeedDownUpsilon} and \ref{FeedDownCharm}.  The vacuum $p+p$ feed-down fractions from LHCb are denoted by solid lines, while our estimates for the in-medium modified feed-down from a purely thermal modification of the quarkonium spectra is given as dashed lines. The fact that all correction factors $\kappa$ are significantly smaller than unity tells us that around the phase transition, thermal fluctuations destabilize the P-wave states efficiently, so that they are not able to contribute to the radiative feed-down to the S-wave states with the same strength as at $T=0$. 

\section{Conclusion}
\label{Sec:Concl}

We have presented a study of the thermal spectral properties of bottomonium and charmonium bound states in the P-wave channel. Our computation is based on a non-relativistic complex potential, recently extracted from first principle lattice QCD simulations.

Similar to our previous investigation of the S-wave channel, we find that thermal fluctuations lead to a sequential melting and a hierarchical broadening of the spectral bound-state features, as well as characteristic shifts to lower energies as shown in Fig.~\ref{Fig:ChrmBotRhoOvervwP}. The larger the vacuum binding energy of a state, the weaker the influence of temperature. In the presence of finite angular momentum an intricate interplay between screening of the formerly confining real-part of the potential and the centrifugal term leads to the dynamical generation of a centrifugal barrier. It appears to contribute to a stabilization of the bound states remnants, which persist as distinct features up to temperatures, where they are already fully engulfed by the continuum.

We provide melting temperatures for the P-wave based on the conventional definition that a bound state is dissolved once its in-medium binding energy equals its thermal width. As expected from the hierarchy of in-medium modifications and in agreement with recent lattice NRQCD studies, we find that the melting temperature of $\chi_b(1S)$  lies very close to the one of the $J/\Psi(1S)$ state, which both share an almost equal vacuum binding energy. 

Phenomenologically the physics of P-wave states provides an important ingredient to the understanding of the production of S-wave quarkonium, due to the possibility of radiative feed-down. Here we computed estimates for the in-medium feed-down contributions based on the feed-down fractions measured recently in $p+p$ by the LHCb collaboration. We find that thermal fluctuations significantly weaken the P-wave states, so that their contribution to feed-down reduces by at least half for the $(1S)$ ground state and even more strongly for the higher lying S-wave states as shown Tab.~\ref{t:feeddowncorrT} and visualized in Fig.~\ref{FeedDownUpsilon} and \ref{FeedDownCharm}.

Even though our study is based on a leading order pNRQCD formulation, it revealed important features of the charmonium and bottomoniun
P-wave states. In a next step towards a more accurate non-relativistic description of quarkonium states, higher order terms in the expansion, such as spin-dependent corrections, need to be included. These corrections may both explain the larger deviation of the vacuum masses seen in our calculation compared to the experimental values tabulated by the PDG and also at finite temperature may lead to corrections to the in-medium modifications observed in this study. It will be the task of future studies to devise an appropriate extraction strategy for the pNRQCD corrections from lattice computations.

\section*{Acknowledgments}

The authors thank G. Aarts, P. Petreczky for stimulating discussions. Part of the calculations to determine the in-medium potential from lattice QCD were performed on the Bielefeld GPU cluster, on the ITP in-house cluster in Heidelberg and the SuperB cluster at EPFL. YB is supported by SNF grant PZ00P2-142524.

\end{document}